\documentclass[prb,showpacs,preprintnumbers,amssymb,twocolumn]{revtex4}
\usepackage{graphicx}% Include figure files
\usepackage{dcolumn}% Align table columns on decimal point
\usepackage{bm}% bold math
\def\be{\begin{equation}}
\def\ee{\end{equation}}
\def\xxm{x_{\rm 2min}}
\def\tK{\tilde{K}_{11}}
\def\ml{\ell}

\begin{document}

\title{Conductance distribution in strongly disordered mesoscopic systems in three dimensions}
\author{K. A. Muttalib$^1$, P.~Marko\v{s}$^2$ and P. W\"olfle$^3$}
\affiliation{$^1$Department of Physics, University of Florida, P.O. Box 118440,
Gainesville, FL 32611-8440\\
$^2$Institute of Physics, Slovak Academy of Sciences - 845 11
Bratislava, Slovakia\\
$^3$Institut f\"ur Theorie der Kondensierten Materie,
Universit\"at Karlsruhe, Germany}
\begin{abstract}

Recent numerical simulations have shown that the distribution of
conductances $P(g)$ in three dimensional strongly localized
systems differs significantly from the expected log normal
distribution. To understand the origin of this difference
analytically, we use a generalized Dorokhov-Mello-Pereyra-Kumar
(DMPK) equation for the joint probability distribution of the
transmission eigenvalues which includes a phenomenological
(disorder and dimensionality dependent) matrix $K$ containing
certain correlations of the transfer matrices.  We first of all
examine the assumptions made in the derivation of the generalized
DMPK equation and find that to a good approximation they remain
valid in three dimensions (3D). We then evaluate the matrix $K$
numerically for various strengths of disorder and various system
sizes. In the strong disorder limit we find that $K$ can be
described by a simple model which, for a cubic system, depends on
a single parameter. We use this phenomenological model to
analytically evaluate the full distribution $P(g)$ for Anderson
insulators in 3D. The analytic results allow us to develop an
intuitive understanding of the entire distribution, which differs
qualitatively from the log-normal distribution of a Q1D wire. We
also show that our method could be applicable in the critical
regime of the Anderson transition.

\end{abstract}

\pacs{73.23.-b, 71.30., 72.10. -d}

\maketitle

\section{Introduction}

The full distribution of conductances $P(g)$ for non-interacting
electrons at zero temperature has recently been studied in detail
in quasi one dimension (Q1D) both analytically \cite{mu-wo-go-03}
and numerically \cite{pichard-nato,q1dnumerics}. Large mesoscopic
fluctuations lead to several remarkable features in the
distribution, including a highly asymmetric `one-sided' log-normal
distribution at intermediate disorder between the metallic and
insulating limits \cite{mu-wo-99,go-mu-wo-02}, and a singularity
in the distribution near the dimensionless conductance $g\sim 1$
in the insulating regime \cite{mu-wo-ga-go-03}. While some
numerical studies exist for 3D finite size systems
\cite{MK-PM,markos1,SMO,markos2,3dnumerics}, there is no analytic
method currently available to study the full distribution $P(g)$
in 3D, especially at strong disorder. Theoretical work based on
$2+\epsilon$ dimensions ($\epsilon \ll 1$), where a weak disorder
approximation can be applied, has been used to propose that the
critical distribution at the Anderson transition point has a
Gaussian center with power law tails \cite{altshuler, shapiro},
but this can not be compared with numerical results in 3D
\cite{MK-PM,markos1} that show a highly non-trivial asymmetric
distribution similar to the one-sided log-normal form of Q1D. It
has not been possible so far to study analytically even the
simpler case of $P(g)$ in the deeply insulating regime in 3D,
where numerical results point to non-trivial deviations from the
expected log-normal form \cite{markos2}.

The $P(g)$ in Q1D systems were studied analytically within the
transfer matrix approach \cite{pichard-nato}. In this paper we use
a recently proposed generalization of the Q1D approach
\cite{mu-go-02} to obtain analytically for the first time the full
$P(g)$ for strongly disordered 3D systems. A brief account of the
work has been published earlier \cite{mmwk}. In Sections II and
III we review briefly the DMPK equation and its generalization,
respectively. In Sect. IV we analyze in detail the numerical data
for 3D disordered systems in all three transport regimes:
metallic, insulating and critical. Numerical data allow us to
determine the free parameters of a matrix $K$ which appears in the
generalized DMPK equation. In Sect V we use them to formulate a
simple model for $K$, and solve the generalized DMPK equation
analytically. In Sect. VI, an analytical formula for the
conductance distribution $P(g)$ is derived in detail.  In our
model, the form of $P(g)$ is determined by two parameters,
$\Gamma$, which measures the strength of the disorder, and
$\gamma_{12}$, which determines the strength of the interaction
term in the generalized DMPK equation. $\gamma_{12}\equiv 1$ in
the Q1D systems.  The fact that $\gamma_{12}<1$ in 3D makes the
statistics of the conductance in 3D different from that in Q1D.
Although we introduced two new disorder dependent parameters, they
turn out to be related to each other and we show that the present
model is not in contradiction with the single parameter scaling
theory \cite{AALR}. In Sect. VII we compare the analytical formula
for the conductance distribution with the numerical data and
analyze how the distribution $P(\ln g)$ depends on the parameter
$\gamma_{12}$. In the limit $\gamma_{12}\to 1$, we recover the Q1D
results. Sect. VIII discusses the possible extension of our
solution to the critical point. We show that our results describe
the critical regime qualitatively correctly, including the non -
analyticity of the critical conductance distribution. Finally,
summary and conclusions are given in Sect. IX.

\section{The transfer matrix approach}

The distribution of conductances for non-interacting electrons at
zero temperature can be studied within the transfer matrix
approach. In this approach, a conductor of length $L_z$ and
cross-section $L\times L$ is placed between two perfect leads; the
scattering states at the Fermi energy then define $N\propto L^2$
channels. The $2N \times 2N$ transfer matrix $M$ relates the flux
amplitudes on the right of the system to those on the left
\cite{muttalib}. Flux conservation and time reversal symmetry (we
consider the case of unbroken time reversal symmetry only)
restricts the number of independent parameters of $M$ to $N(2N+1)$
and $M$ can be written in general as \cite{dmpk,muttalib}
\begin{equation}\label{one}
M=\left(\matrix{ u & 0  \cr 0 & {u}^* \cr }\right) \left(\matrix{
\sqrt{1+\lambda} & \sqrt{\lambda}   \cr \sqrt{\lambda}   &
\sqrt{1+\lambda} \cr }\right)\left(\matrix{ v & 0  \cr 0 & {v}^*
\cr }\right),
\end{equation}
where $u,v$ are $N \times N$ unitary matrices, and $\lambda$ is a
diagonal matrix, with positive elements $\lambda_i, i=1,2, ...N$.
Microscopic distribution of impurities will lead to a distribution
$p_{L_z}(M)d\mu (M)$ of the transfer matrices where $d\mu(M)$ is an
invariant measure which we rewrite as
\begin{equation}
p_{L_z}(M)d\mu (M) = p_{L_z}(\lambda,u,v)d\mu (\lambda)d\mu (u)d\mu (v).
\end{equation}
If we know the marginal distribution
\begin{equation}
\bar{p}_{L_z}(\{\lambda_a\}) = \int p_{L_z}(\lambda,u,v)d\mu (u)d\mu (v),
\end{equation}
then the distribution of conductances $P(g)$ can be written as
\begin{equation}
P(g) = \int \cdots \int \prod_{a=1}^N d\lambda_a
\bar{p}_{L_z}(\{\lambda_a\}) \delta\left(g-\sum_{a=1}^N
\frac{1}{1+\lambda_a}\right),
\end{equation}
where
\begin{equation}\label{eq-landauer}
g=\sum_{a=1}^N \frac{1}{1+\lambda_a}
\end{equation}
is the Landauer conductance \cite{landauer}. A systematic approach
to evaluate the $N$-dimensional integral, based on a mapping to a
one-dimensional statistical mechanical problem, has been developed
\cite{mu-wo-go-03}, so the full distribution $P(g)$ can be
obtained if the marginal distribution
$\bar{p}_{L_z}(\{\lambda_a\})$ is known. Note that the
distribution of other transport variables which can be written as
$\sum_a f(\{\lambda_a\})$, e.g. shot noise power \cite{blanter}
$P=\sum_{a=1}^N \frac{\lambda_a}{(1+\lambda_a)^2}$ or conductance
of N-S (Normal metal-Superconductor) microbridge
\cite{beenakkerNS} $G=\sum_{a=1}^N \frac{1}{(1+2\lambda_a)^2}$,
can also be obtained in the same way. The above approach is valid
in principle for all strengths of disorder, in all dimensions.

If we assume that the distribution $p_{L_z}(\lambda,u,v)$ is {\it
independent} of $u,v$, then the evolution of the distribution with
length $L_z$ can be obtained from a Fokker-Planck equation first
derived by Dorokhov and by Mello, Pereyra and Kumar \cite{dmpk}
which has become known as the DMPK equation:
\begin{eqnarray}\label{DMPK}
\frac{\partial p_{L_z}(\lambda)}{\partial (L_z/\ml)}
&=&\frac{2}{N+1}\frac{1}{J}\sum_a^N
\frac{\partial}{\partial\lambda_a}\left[\lambda_a(1+\lambda_a)
J\frac{\partial p}{\partial \lambda_a}\right],\cr J & \equiv &
\prod_{a<b}^N|\lambda_a-\lambda_b|^{\beta}.
\end{eqnarray}
Here $\ml$ is the mean free path and 
the parameter $\beta$ is equal to $1,2$ or $4$ depending on
orthogonal, unitary or symplectic symmetry of the transfer
matrices. We will consider only the case with time-reversal
symmetry, for which $\beta=1$. Although the parameters $\lambda_a$
are not eigenvalues of $M$, it turns out that they determine the
eigenvalues of the matrix $TT^\dag$ ($T$ is the transmission
matrix\cite{Pth}) \be \label{tt}  TT^\dag=v^*(1+\lambda)^{-1}v \ee
which characterizes the conductance given by
Eq.~(\ref{eq-landauer}), and the matrix $v$ contains the
eigenvectors of $TT^\dag$. So we will loosely refer to these as
the eigenvalues and the eigenvectors in the text. Note that the
parameter $\beta$ determines the strength of `level repulsion'
between eigenvalues.

The assumption that $p_{L_z}(\lambda,u,v)$ is independent of $u,v$
restricts the validity of the DMPK equation to quasi one dimension
(Q1D). Quasi 1D means not only that $L_z \gg L$ where $L_z$ is the
direction of the current and $L$ is the cross-sectional dimension,
but it also requires that $\xi \gg L$, where $\xi$ is the
localization length. In this limit, all channels become
`equivalent', the matrices $u$ and $v$ become isotropic and the
distribution becomes independent of $u$ or $v$. The distribution
of conductances $P(g)$ for such Q1D systems  has been studied
recently; it has many surprising features arising from large
mesoscopic fluctuations. These include a highly asymmetric
`one-sided' log-normal distribution at intermediate disorder
between the metallic and insulating limits
\cite{mu-wo-99,go-mu-wo-02}, and a singularity in the distribution
near $g\sim 1$ in the insulating regime \cite{mu-wo-ga-go-03}. It
is not clear if these features persist in higher dimensions.

\section{Generalized DMPK equation in higher dimensions}

To study 3D systems, a phenomenological generalization of the DMPK
equation has recently been proposed in which the
Q1D
restriction is lifted in favor of an unknown
matrix
\begin{equation}\label{two}
K_{ab}\equiv \left<k_{ab}\right>_L; \;\;\;  k_{ab} \equiv
\sum_{\alpha=1}^N|v_{a\alpha}|^2|v_{b\alpha}|^2,
\end{equation}
where the angular bracket represents an ensemble average. In terms
of this matrix, the marginal distribution $\bar{p}_{L_z}(\lambda)$
satisfies an evolution equation given by \cite{mu-go-02}
\begin{eqnarray}\label{three}
\frac{\partial \bar{p}_{L_z}(\lambda)}{\partial (L_z/\ml)}
&=&\frac{1}{\bar{J}}\sum_a^N
\frac{\partial}{\partial\lambda_a}\left[\lambda_a(1+\lambda_a)K_{aa}
\bar{J}\frac{\partial \bar{p}}{\partial \lambda_a}\right],\cr
\bar{J}&\equiv&\prod_{a<b}^N|\lambda_a-\lambda_b|^{\gamma_{ab}};
\;\;\; \gamma_{ab}\equiv\frac{2K_{ab}}{K_{aa}}.
\end{eqnarray}
In Q1D under the isotropy condition, the matrix $K$ reduces to
\begin{equation}\label{dmpk-q1d}
K^{Q1D}_{ab}=\frac{1+\delta_{ab}}{N+1}; \;\;\;
\gamma^{Q1D}_{ab}=1,
\end{equation}
and one recovers the DMPK equation (with $\beta=1$). In 3D, $K$ is
not known analytically, and must be obtained from independent
numerical studies.

There are two major assumptions made in
Ref.~[\onlinecite{mu-go-02}] in deriving Eq.~(\ref{three}):
\begin{itemize}
\item[(i)]  the elements $k_{ab}$ can be
replaced by their mean values $K_{ab}$, and
\item[(ii)] the $L_z$-dependence of $K_{ab}$ is negligible.
\end{itemize}
These assumptions need to be verified before the equation can be
used. Note that the matrix $K$ depends on the choice of
representation. Since the assumptions are most natural in the
position representation, we will study the matrix in this
representation.

%%%%%%%%%%%%%%%%%%%%%%%%%%%%%%%%%%%%%%%%%%%%%%%%%%%%%%%%%%%%%%%%%%%%%%%%%%%%%%%%%%%%%%%%%%%%%5

\section{Numerical data}\label{numerical:data}

The generalized DMPK equation apparently introduces a large number
of new parameters, elements of the matrix $K_{ab}$. There is no
theoretical prediction about how these parameters should depend on
the size of the system or on disorder. We only know that in the
Q1D limit they should follow Eq.~(\ref{dmpk-q1d}). Therefore our
first goal is to study numerically various 3D and Q1D systems
systematically in detail in order to answer the following
questions:

\begin{itemize}
\item[\textbf{Q1:}] Are the assumptions (i) and (ii) discussed in Section III valid
at all strengths of disorder?
\item[\textbf{Q2:}] How do the elements $K_{ab}$ depend on disorder and on the system size?
\item[\textbf{Q3:}] How do the elements $K_{ab}$ depend on the indices $a$ and $b$?
\item[\textbf{Q4:}] Given the size, disorder and index dependence of
$K_{ab}$, is it possible to construct a simple model of $K$ at all
disorder with only a small number of parameters?
\end{itemize}
We will address all of the above in this section, but let us first
briefly discuss the numerical procedure used to evaluate $K$.

We consider the tight binding Anderson model defined by the
Hamiltonian
\begin{equation}\label{AndHam}
{\cal H}=W\sum_n \varepsilon_nc_n^\dag c_n+\sum_{[nn']}
t_{nn'}c^\dag_n c_{n'}.
\end{equation}
In Eq.~(\ref{AndHam}), $n=(xyz)$ counts sites on the simple cubic
lattice of the size $L\times L\times L_z$, and $\varepsilon_n$ are
random energies, uniformly distributed in the interval
$[-\frac{1}{2},\frac{1}{2}]$. The parameter $W$ measures the
strength of disorder. The Fermi energy is chosen as $E_F=0.01$.
The hopping term $t_{nn'}$ between the nearest-neighbor sites
$nn'$ is unity for hopping along the $z$ direction and $t_{nn'}=t$
for hopping in the $x$ and $y$ directions.  Then the dispersion
relation is \be\label{ef} E=2\cos k_z +2t\cos k_x + 2t\cos k_y \ee
For a given cross section of the sample: $L_x=L_y=L$, $k_x$ and
$k_y$ possess values $\pi/(L+1)\times 1,2,\dots L$ (we consider
hard wall boundary conditions). At fixed energy E,  given values
of $k_x$ and $k_y$ determine the value of $k_z$, which is either
real (if $|\cos k_z|<1$) or imaginary. The latter case corresponds
to closed channels which do not transmit current in perfect leads.
To avoid these closed channels, which are not considered in the
DMPK formulation, we use $t=0.4$.  Then the model
Eq.~(\ref{AndHam}) exhibits a metal-insulator transition
\cite{zambetaki} at $W_c\approx 9$. To obtain transport
properties, we use the transfer matrix  method of
[\onlinecite{pendry}]. The main difference from previous works
\cite{SMO,markos2} is that we also calculate eigenvectors of the
matrix $TT^\dag$. Using
Eq.~(\ref{one}), we calculate numerically the matrix $TT^\dag$.
Owing to Eq.~(\ref{tt}), diagonalizing $TT^\dag$ gives us
$\lambda$ as well as all elements of the matrix $v$.

Note that the eigenvectors depend on the representation. In the
original formulation of the DMPK approach, semi-infinite leads
consist of mutually independent and equivalent 1D wires.
Therefore, the transfer matrix in the leads is diagonal in both
channel and space representations. In numerical work, we need to
distinguish between these two, since the transfer matrix is
diagonal only in the channel representation. We therefore
calculate the matrix $TT^\dag$ in the channel representation, find
eigenvalues and eigenvectors, and transform the latter back to the
space representation to obtain the matrix $v$. Elements of $v$ are
then used for the calculation of the matrix $K$ in the space
representation.

We now go back and address the questions raised at the beginning
of this section.

\subsection{Q1: Validity of the assumptions (i) and (ii)}

\begin{figure}[t]
\includegraphics[clip,width=0.35\textwidth]{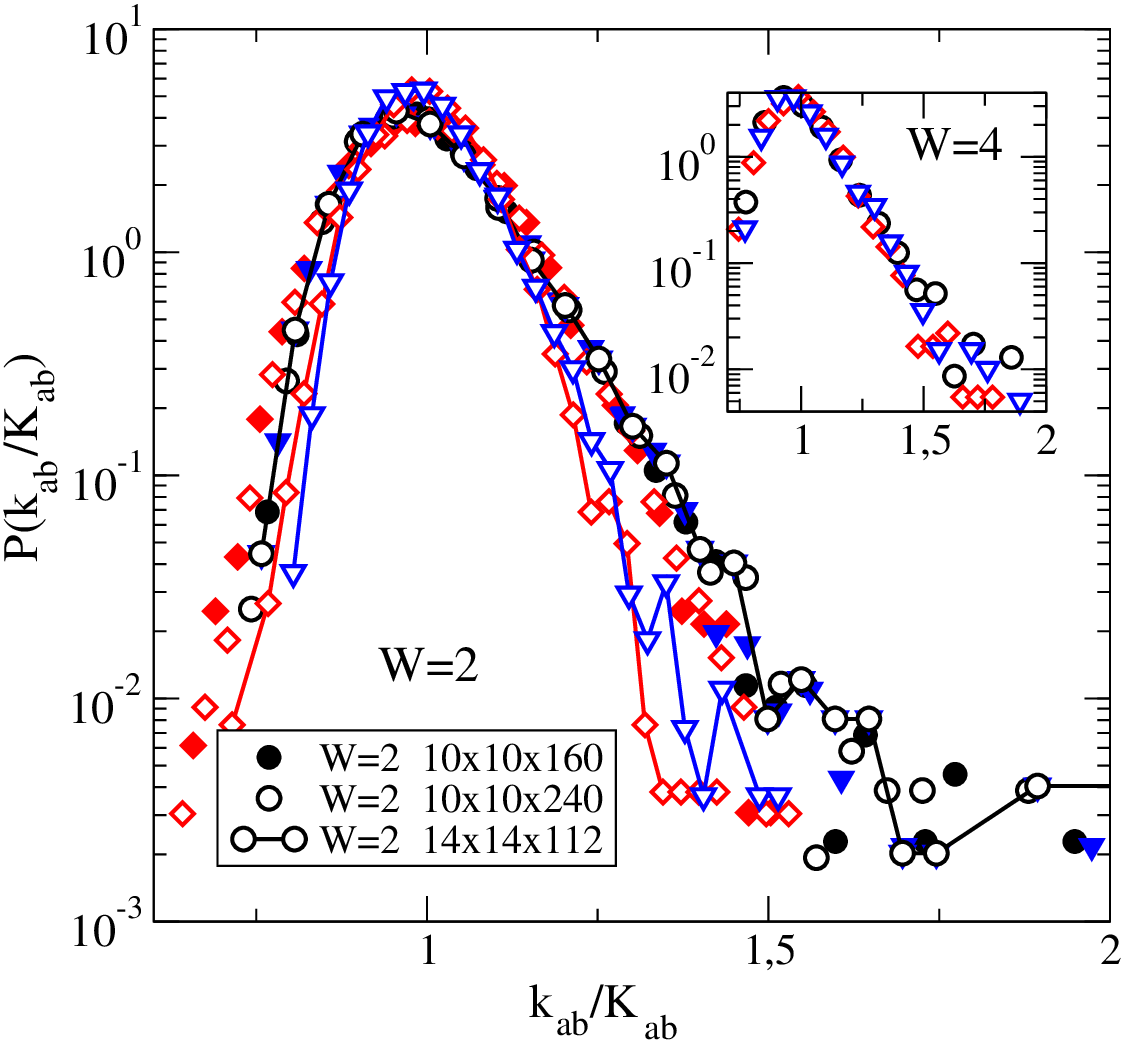}
\caption{Probability distribution of \textsl{normalized} $k_{ab}$
for Q1D systems. ($\circ$: $a=b=1$, $\diamondsuit$: $a=1$, $b=2$,
$\bigtriangledown$: $a=b=2$). Full symbols: $W=2$, $L=10$,
$L_z=16L$; open symbols: $W=2$, $L=10$, $L_z=24L$. Lines with
symbols:  $W=2$, $L=14$, $L_z=8L$. Data confirm that the
distribution becomes narrower when $L$ increases. Inset shows the
same for $W=4$ and $L=18$, $L_z=4L$. As expected, distributions
are broader. } \label{3D_t04_Q1D_k}
\end{figure}

\begin{figure}[t]
\includegraphics[clip,width=0.4\textwidth]{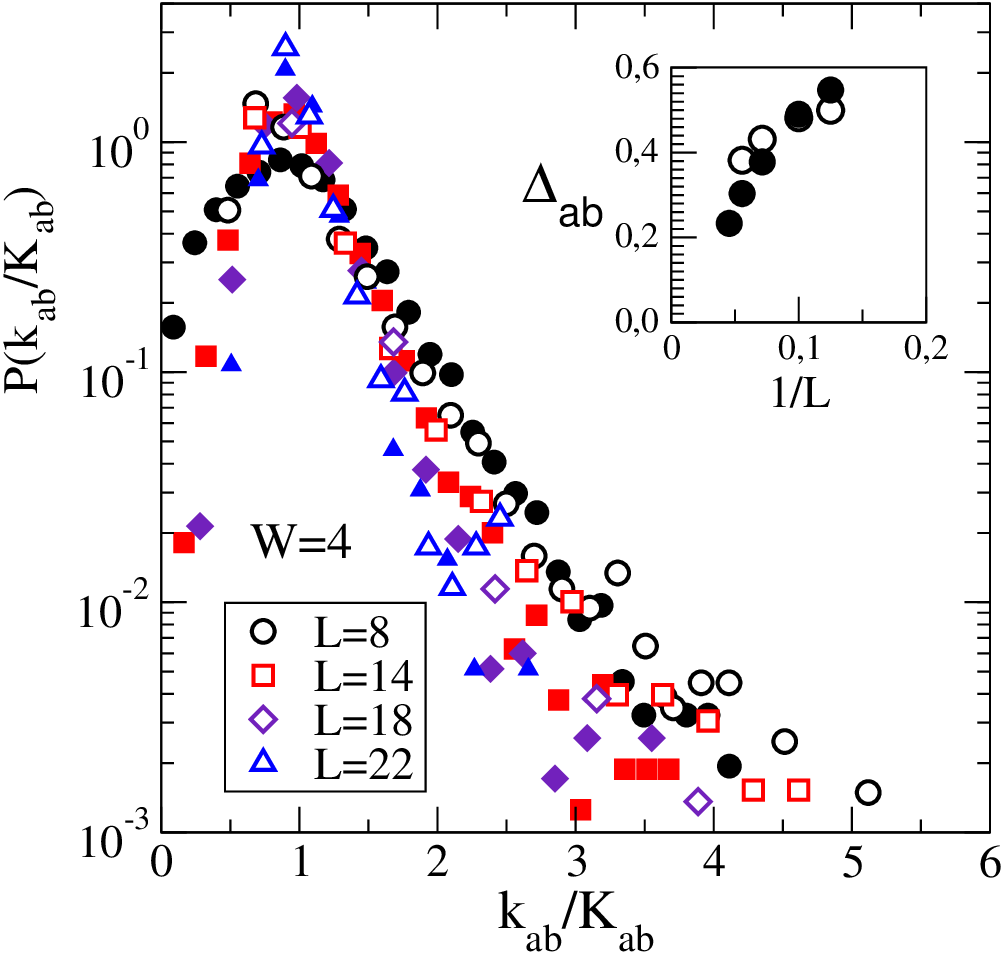}~~
\caption{The distribution of \textsl{normalized} $k_{ab}$ for 3D
systems with $W=4$ and various system sizes. (critical disorder
$W_c\approx 9$.) Open symbols: $a=b=1$; full symbols: $a=1$,
$b=2$. Inset shows the  $L$-dependence of
$\Delta_{ab}=\sqrt{\textrm{var}~k_{ab}}/K_{ab}$ which decreases
when $L$ increases. Data confirm that distribution is
self-averaging (it becomes narrower and $\Delta_{ab}\to 0$ when
$L$ increases) although  it is broader than in the Q1D case.}
\label{3D_t04_kov_k}
\end{figure}

\begin{figure}
\includegraphics[clip,width=0.4\textwidth]{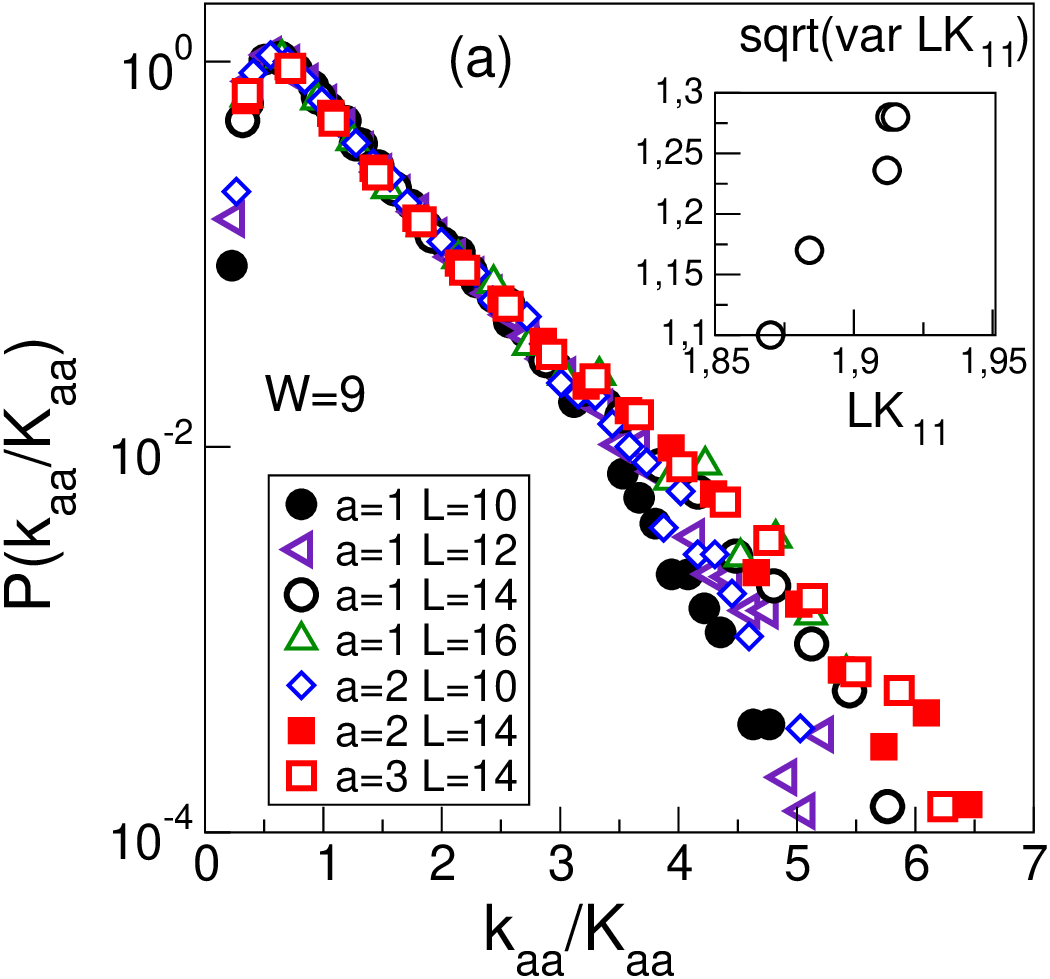}
\includegraphics[clip,width=0.4\textwidth]{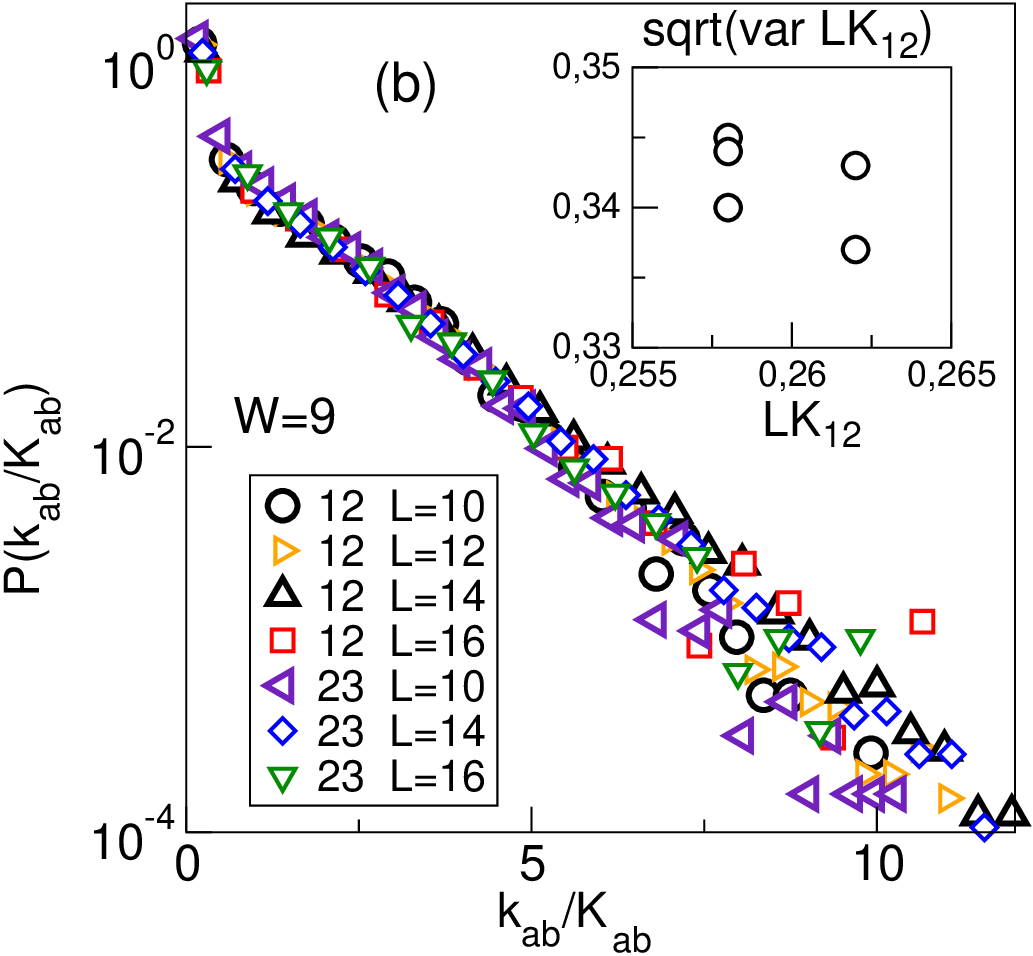}
\caption{Probability distribution of \textsl{normalized} matrix
elements $k_{ab}$. (a) $P(k_{aa}/K_{aa})$ ($a=1,2$ and 3) at the
critical point  $W=W_c\approx 9$. Distribution is $L$-independent
(apart from the exponential tail which is broader for larger $L$
since mean value $K_{aa}\sim 1/L$). Note the logarithmic scale on
the $y$ axis. (b) Distribution of off-diagonal elements $k_{12}$
and $k_{23}$ possess sharp maxima close to zero, and long
exponential tails. Insets show standard deviations of
distributions as a function of mean values for $8\le L\le 18$.
These data also show the accuracy of our estimate of the critical
point since we expect both the mean and the standard deviation to
be $L$-independent at $W=W_c$. The distributions for Q1D systems
$L\times L\times 8L$ are almost identical to those for cubes (data
not shown).
%but available: 3D_t04_w95_Q1D.eps
} \label{pkaa_cp}
\end{figure}

\begin{figure}[t]
\includegraphics[clip,width=0.4\textwidth]{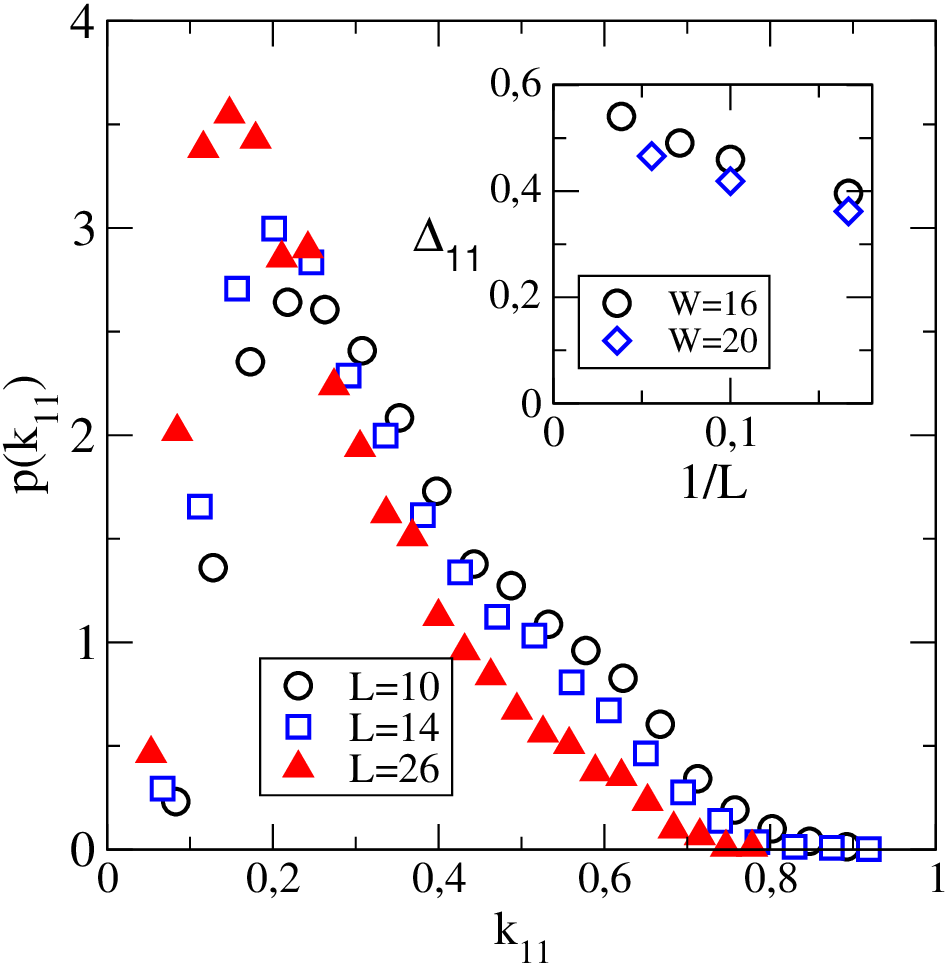}
\caption{Probability distribution $P(k_{11})$ in the insulating
regime ($W=16$). Although distribution becomes narrower when
system size $L$ increases, it is not self-averaging. Inset shows
the size dependence of the $\Delta_{11}=\sqrt{\textrm{var}~k_{11}}/K_{11}$
for $W=16$ and $W=20$. $\Delta_{11}$  converges to a non-zero
constant when $L\to\infty$. } \label{loc_a}
\end{figure}

In order to check assumption (i), i.e. if the elements $k_{ab}$
can be replaced by their average $K_{ab}$ in 3D, we analyze the
probability distribution $P(k_{ab})$. We start with the weakly
disordered metallic regime. Two values of disorder were used:
$W=2$ and $W=4$. More relevant than the actual strength of the
disorder is the mean free path $\ml$ which can be estimated from
the mean conductance \cite{pichard} \be\label{mfp} \langle
g\rangle = \frac{N}{L_z}\ml \ee with $N=L^2$. From the $L_z$
dependence of $\langle g\rangle$ in Q1D systems we estimate
$\ml(W=2)\approx 9.2$ and $\ml(W=4)\approx 1.8$, in units of the
lattice spacing.

\begin{figure}
\includegraphics[clip,width=0.35\textwidth]{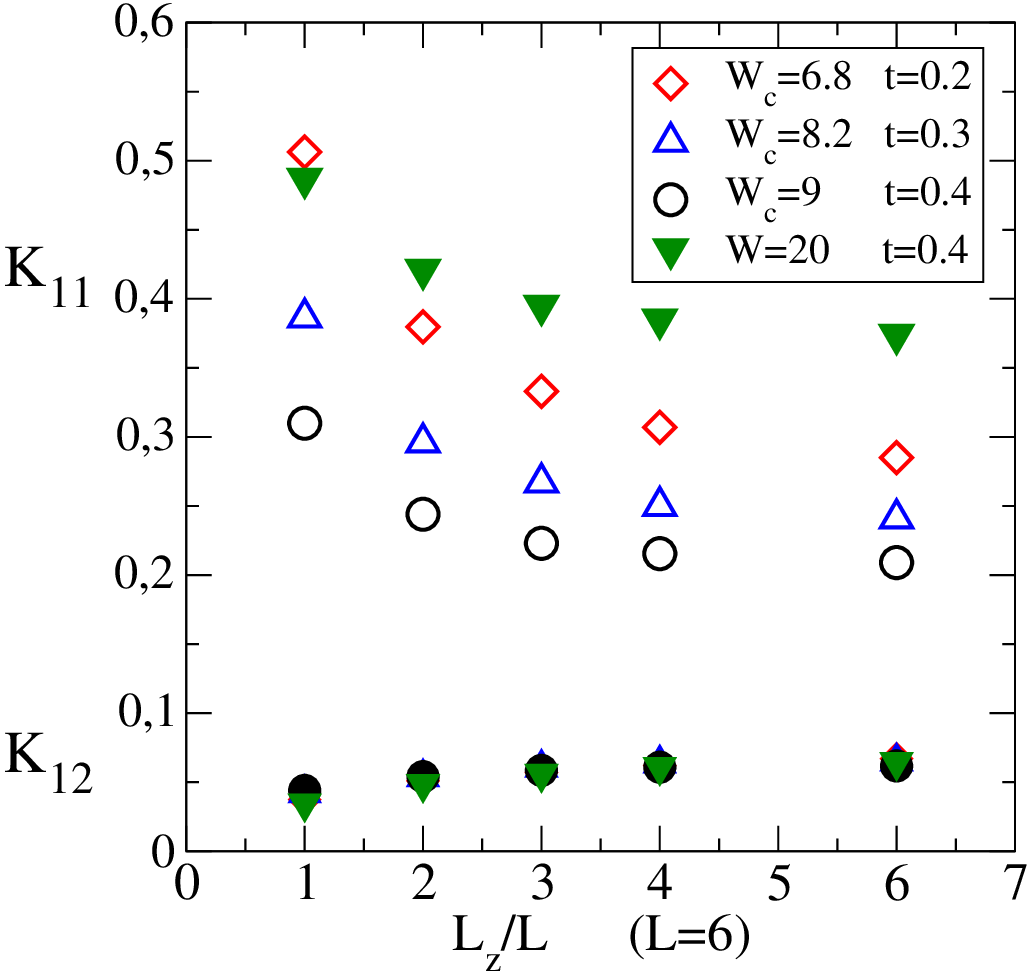}
\caption{%
At the critical point, which depends on the anisotropy parameter
$t$, both $LK_{11}$ and $LK_{12}$ converge to $L_z$-independent
values when $L_z/L\to\infty$. The limiting value $LK_{11}$ depends
on $t$. Shown are also data for the insulating regime $W=20$,
$t=0.4$. All data for $K_{12}$ almost coincide so that they are
not distinguishable in the figure. } \label{3D_t_cp}
\end{figure}

First, we test how assumption (i) is fulfilled in the weakly
disordered limit, where the DMPK equation is known to describe the
universal features of transport statistics correctly. We show in
Figure \ref{3D_t04_Q1D_k} the distribution $P(k_{ab})$ in the
weakly disordered Q1D regime. As expected, the width of the
distribution increases when $W$ increases, but $P(k_{ab})$ is
self-averaging (it becomes narrower when $L$ increases). Figure
\ref{3D_t04_kov_k} shows the same distribution for 3D systems. The
distribution is again  self-averaging, although much broader than
in the Q1D case.

In the critical
regime, the probability distribution $P(k_{ab})$ is no longer
self-averaging but tends to be $L$-independent in the limit
$L\to\infty$ (Fig.~\ref{pkaa_cp}).  Although the distributions possess long
exponential tails, they have well defined sharp maxima, which do
not depend on the system size.

In the insulating regime, the distribution $P(k_{11})$ becomes
narrower when $L$ increases  (Fig.~\ref{loc_a}). However, on the
basis of our numerical data we conclude that the distribution is
not self-averaging. Although var $k_{11}$ decreases when
$L\to\infty$ (data not shown), the normalized width
$\Delta_{11}=\sqrt{\textrm{var}~k_{11}}/K_{11}$ (shown in inset of fig.
\ref{loc_a}) slightly increases when $L$ increases. As $K_{11}$
itself is non-zero in the limit $L\to\infty$ (fig.
\ref{3D_t04_k11}), $\Delta_{11}$ should converge to an
$L$-independent function for large $L$.

The
distribution of off-diagonal elements $P(k_{12})$ in the
insulating regime (not shown) is qualitatively the same as that at
the critical point.

We conclude that both in the critical and localized regimes the
distributions converge to $L$ independent functions with a
well-defined peak, but the standard deviation is of the same order
of magnitude as the mean. We note that the most-probable value of
$k_{11}$ is always very close to its mean value; we therefore
expect that replacing $k_{11}$ by its mean value $K_{11}$ is a
reasonable approximation as long as one is interested in
qualitative results only. Thus, although to leading order
assumption (i) remains valid for all disorder, we have to keep in
mind that fluctuations of the elements $k_{ab}$ in the strongly
disordered regime might become important if the final results are
sensitive to the exact values of these elements. We have checked 
that the final distribution of conductances do not change in any
appreciable way if fluctuations of $k_{11}$ are included by overaging
the conductance distribution over $P(k_{11})$.

To check assumption (ii), namely if the $L_z$ dependence of
$K_{ab}$ is negligible, we studied the $L_z/L$ dependence of
$K_{11}$ and $K_{12}$. Figure \ref{3D_t_cp} confirms that for both
critical and insulating regimes the parameters $K_{11}$ and
$K_{12}$ converge to non-zero (although $t$ - dependent) limits
when the length of the system increases. It shows that the
properties of the matrix $K$ depend only slightly on the ratio
$L_z/L$  and reach $L_z$-independent limiting values when
$L_z/L\to\infty$ in all transport regimes. The  assumption (ii) is
therefore reasonably well satisfied at all disorder as long as
$L_z \ge L$.

Thus we conclude from our numerical studies that to leading
approximation the generalized DMPK equation (\ref{three}) remains
qualitatively valid at all disorder in 3D, but the effect of
fluctuations of $k_{ab}$ on the final results has to be evaluated
in more detail before a quantitative comparison with numerical
results can be made.

\subsection{Q2: Disorder and size dependence of $K_{ab}$}

\begin{figure}[t]
\includegraphics[clip,width=0.35\textwidth]{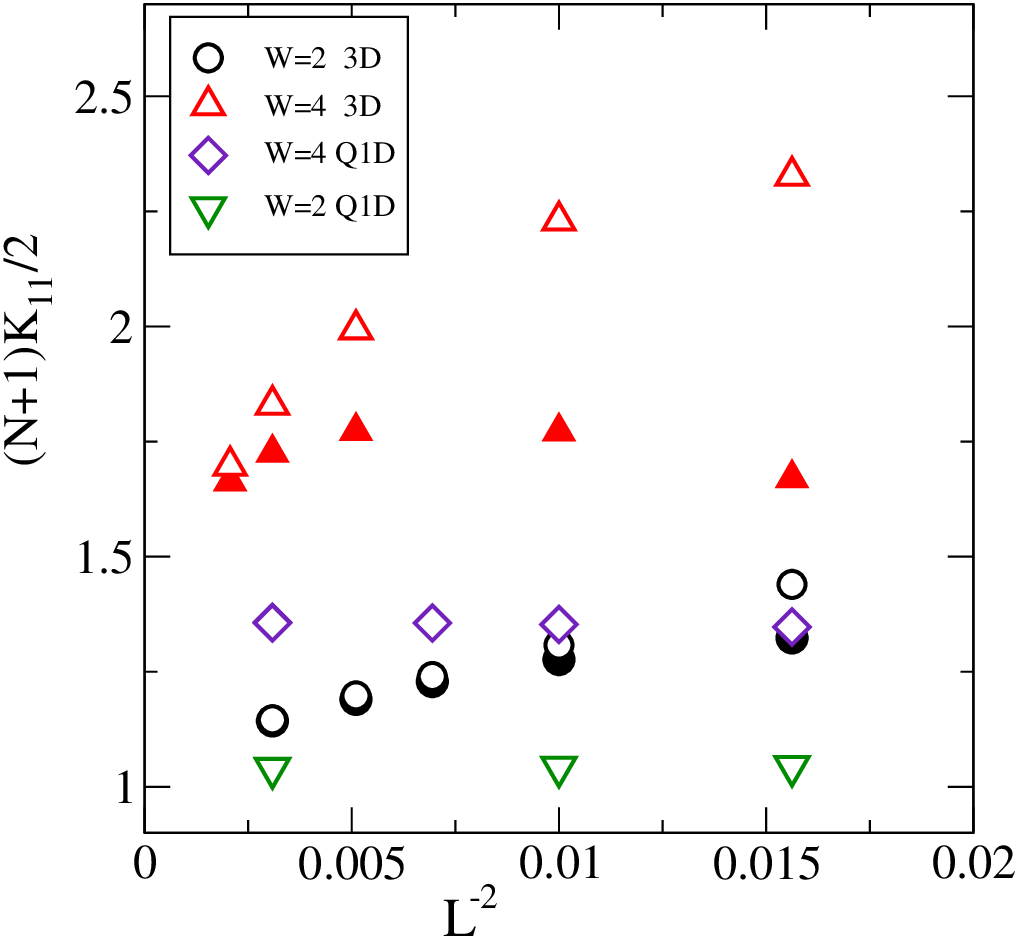}
\caption{$1/L^2$ dependence of $\kappa=(N+1)K_{11}/2$ (open
symbols) and $(N+1)K_{12}$ (full symbols) in  the metallic regime.
Data confirms that $\kappa$ depends on  $L$. This agrees with data
in Fig.~\ref{metal_kaa}. Also, $2K_{12}$ differs from $K_{11}$ for
small $L$. This agrees with data in Fig.~\ref{metal_gamma1a}. To
estimate limiting behavior of $K_{11}$, we also considered Q1D
systems $L\times L\times 4L$. $\kappa$ is close to 1 only for very
weak disorder $W=2$. For $W=4$ we obtained $\kappa\approx 1.36$.
As this value does not depend on $L$ for $8\le L\le 18$
($\Diamond$), we expect that 3D data for $W=4$ will converge to
the same value when $L\to\infty$. } \label{3D_t04_k11_metal}
\end{figure}

We start with the weak disorder regime. To distinguish the generic
$W$ dependence of $K_{11}$ from finite size effects, we analyzed
in Fig.~\ref{3D_t04_k11_metal} the $L$ dependence of the parameter
\be \kappa=(N+1)K_{11}/2. \ee As expected, $\kappa$ decreases when
$L$ increases. However, from the analysis of Q1D systems (also
shown in Fig.~\ref{3D_t04_k11_metal}) we conclude that $\kappa$
converges to 1 only for very small values of disorder. As shown in
Fig.~\ref{3D_t04_k11_metal}, $\kappa\approx 1.36$ for $W=4$. Thus,
deviations from Eq.~(\ref{dmpk-q1d}) already appear in the
metallic limit, probably due to the decrease of the mean free
path.

%\begin{figure}
%\includegraphics[clip,width=0.3\textwidth]{3D_t04_4_18_Lz.eps}
%\caption{$a$ - dependence of $K_{1a}$ for cube and for Q1D system.}
%\label{metal_k1a}
%\end{figure}

%\includegraphics[clip,width=0.3\textwidth]{3D_t04_xa.eps}

\begin{figure}[t]
\includegraphics[clip,width=0.35\textwidth]{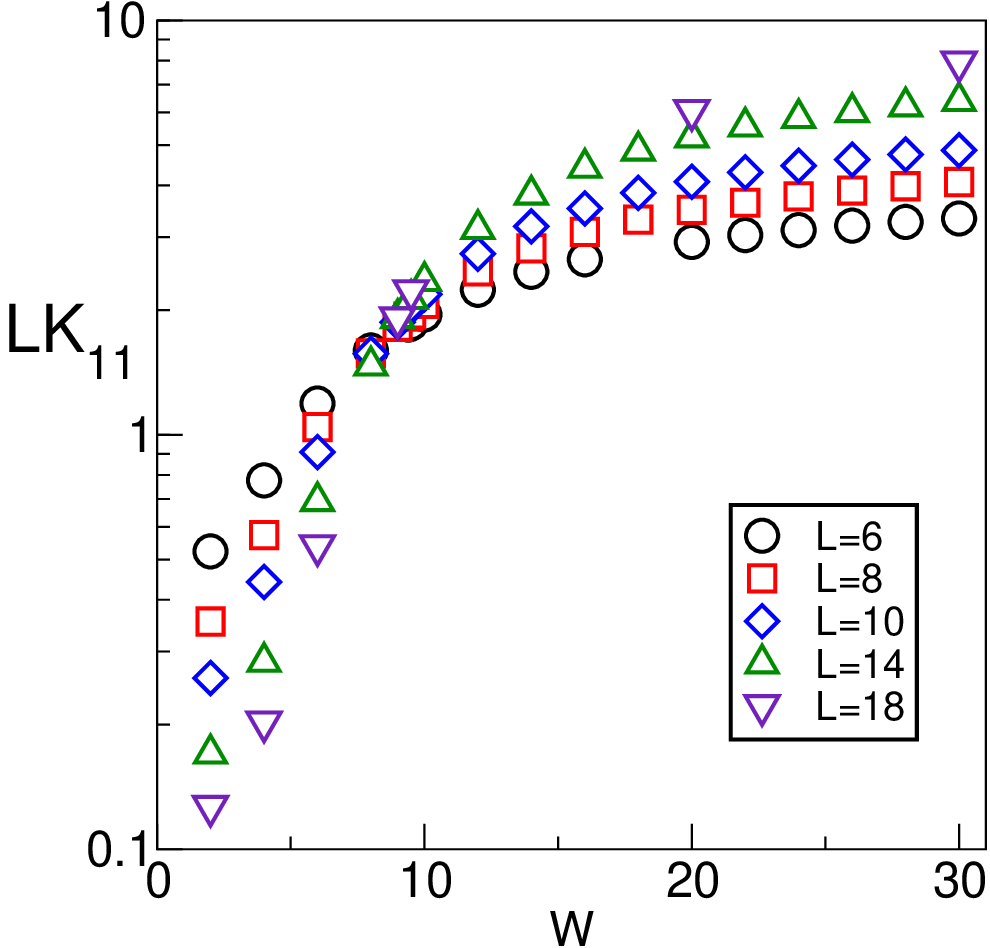}
\includegraphics[clip,width=0.35\textwidth]{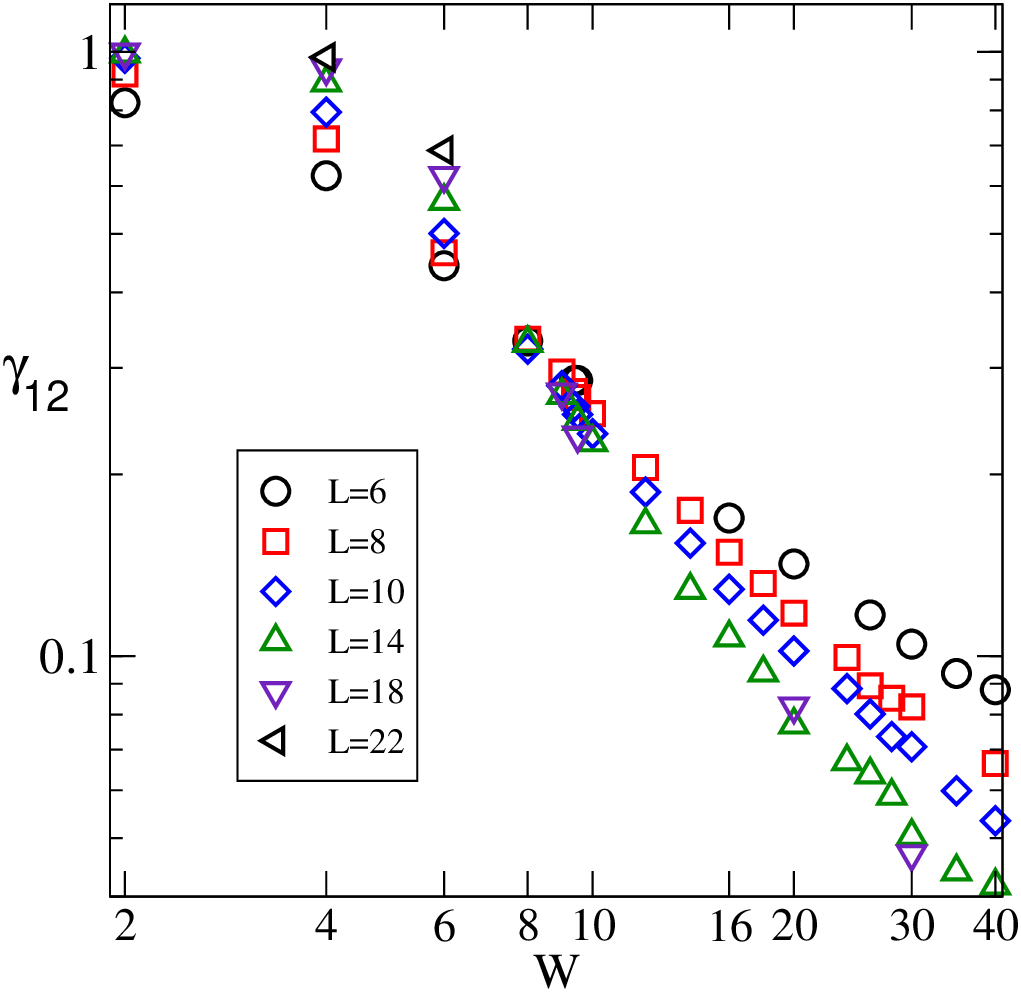}
\caption{Disorder dependence of $K_{11}$ and $\gamma_{12}$ for
various system sizes. Note the common crossing point at
$W=W_c\approx 9$. Note also that $\gamma_{12}\to 1$ for $W<W_c$
and $L\to\infty$, as expected from DMPK, but $\gamma_{12}$
decreases with the system size for $W>W_c$.} \label{W_dependence}
\end{figure}

\begin{figure}
\includegraphics[clip,width=0.35\textwidth]{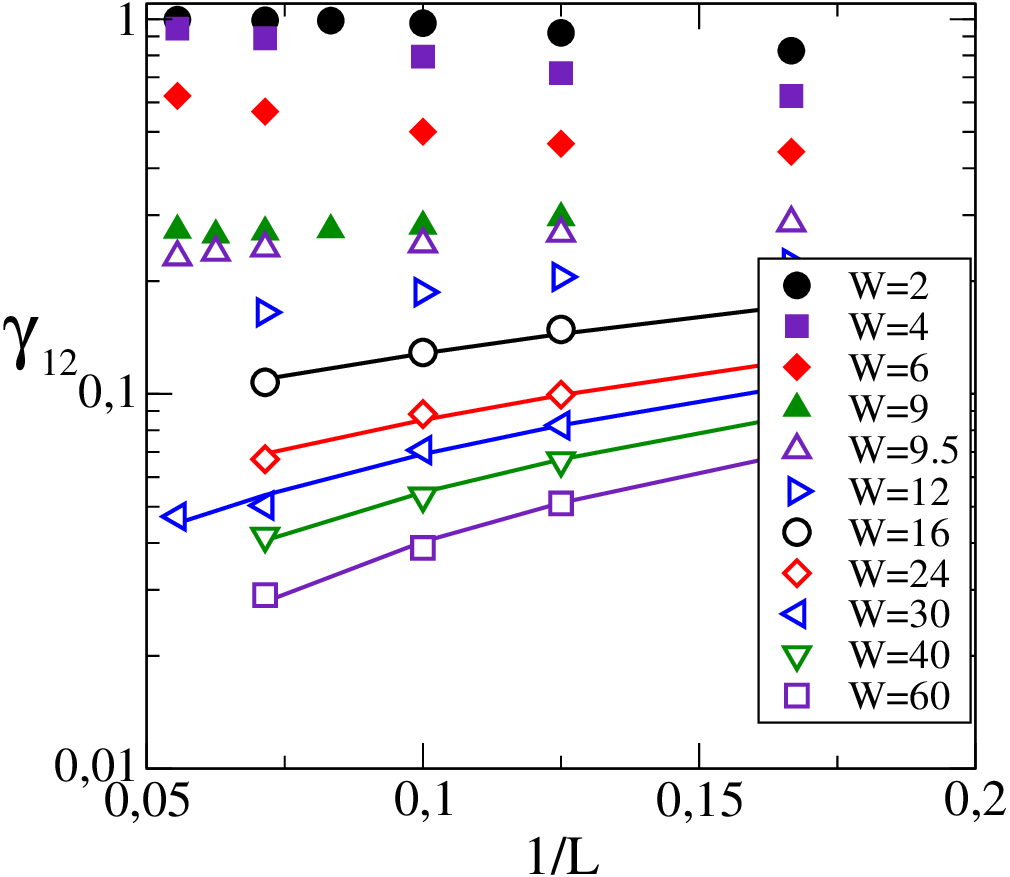}
\caption{$\gamma_{12}$ as a function of the system size for
various strengths of disorder. In the metallic regime,
$\gamma_{12}$ converges to 1 when $L\to\infty$. At the critical
point, $\gamma_{12}$ is $L$-independent, and in the insulating
regime $\gamma_{12}$ decreases when $L$ increases. Solid lines are
linear fits $\gamma_{12}=a+b/L$ with $a\sim 10^{-3}$ for $W\le
40$. } \label{loc_gamma}
\end{figure}

\begin{figure}
\includegraphics[clip,width=0.4\textwidth]{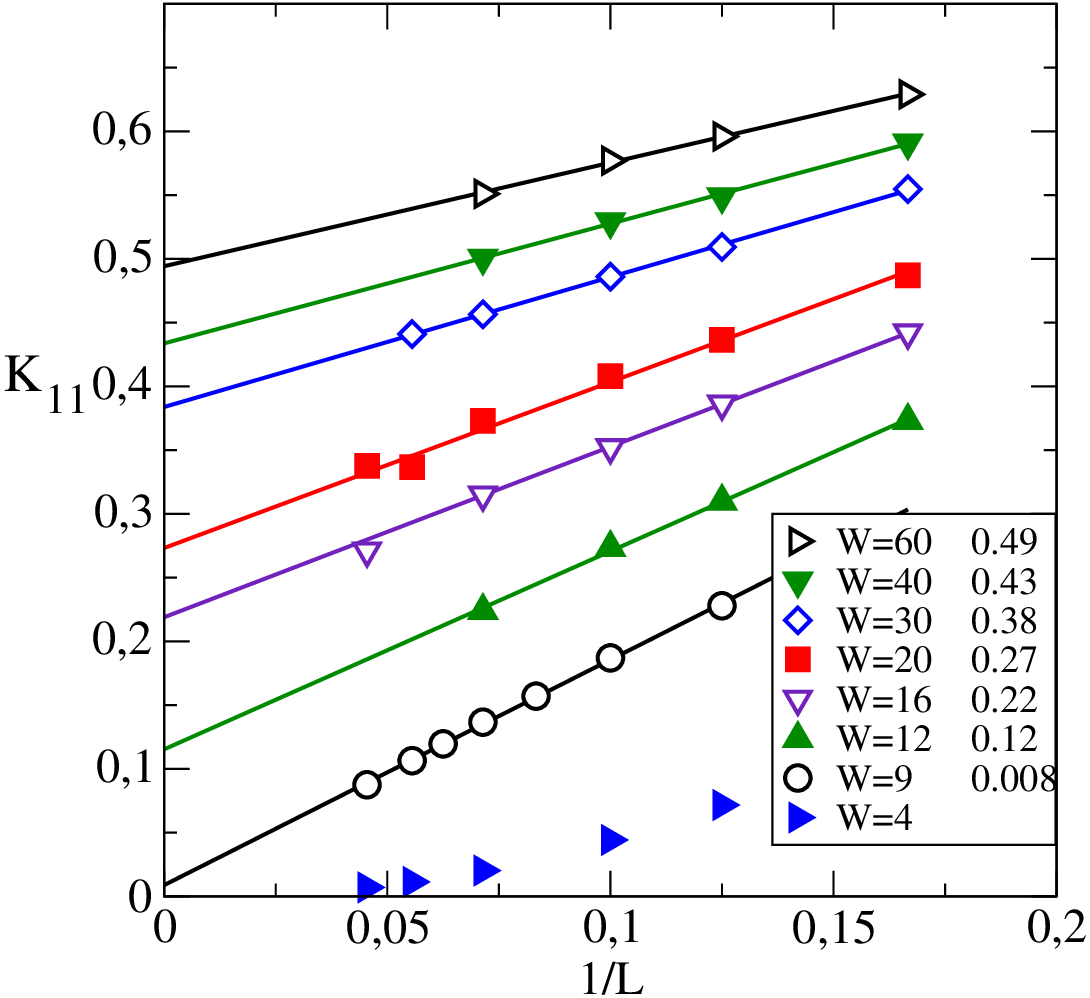}
\caption{System size dependence of $K_{11}$ for various values of
disorder. Solid lines are linear fits. Data in legend presents
limiting values $\tK=\lim_{L\to\infty} K_{11}(L)$. Shown are also
data for the metallic regime ($W=4$) for which $K_{11}\sim 1/L^2$
(Table I).  While $\tK=0$ in metal and at the critical point
($W=9$), it  is non-zero in the insulating regime. Thus, it may be
used as the order parameter of the Anderson transition. }
\label{3D_t04_k11}
\end{figure}

\begin{figure}[t]
\includegraphics[clip,width=0.4\textwidth]{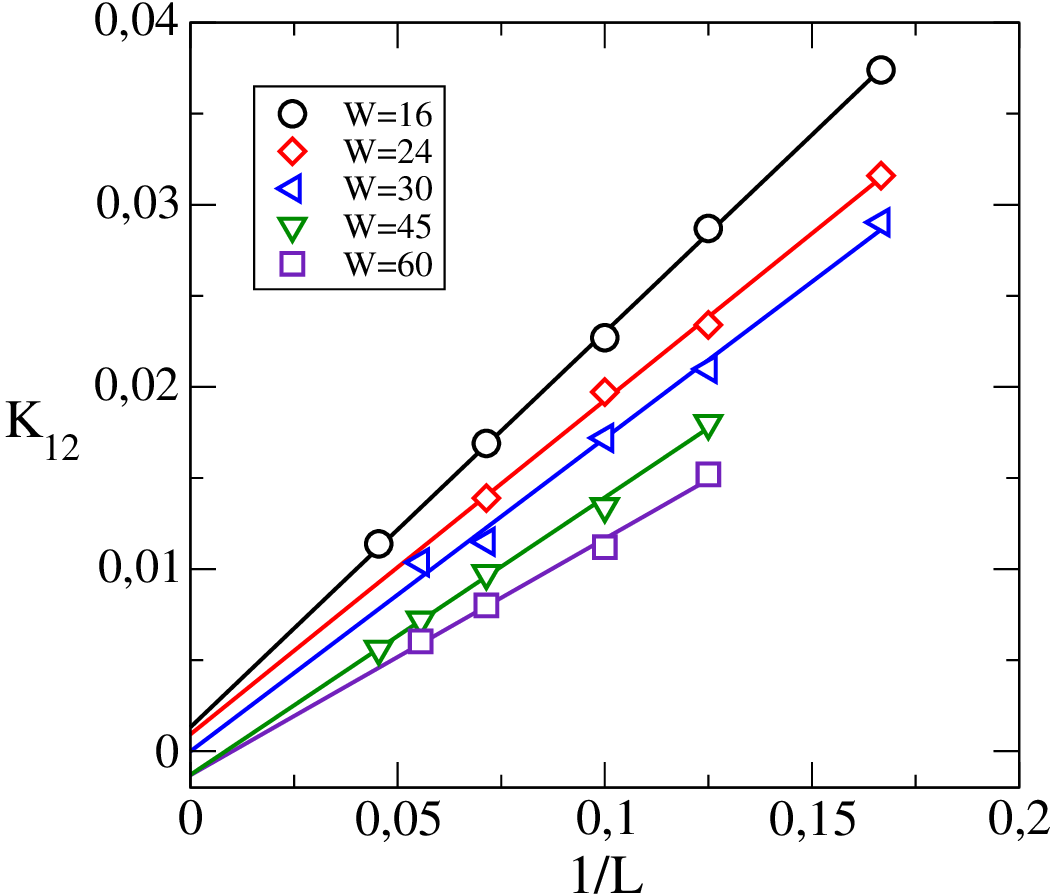}
\caption{System size dependence of $K_{12}$ for various values of
disorder in the insulating regime.  Data  confirm that $K_{12}\sim 1/L$
and that $\lim_{L\to\infty} K_{12}=0$.
}\label{3D_t04_k12}
\end{figure}

\begin{figure}[h]
\includegraphics[clip,width=0.35\textwidth]{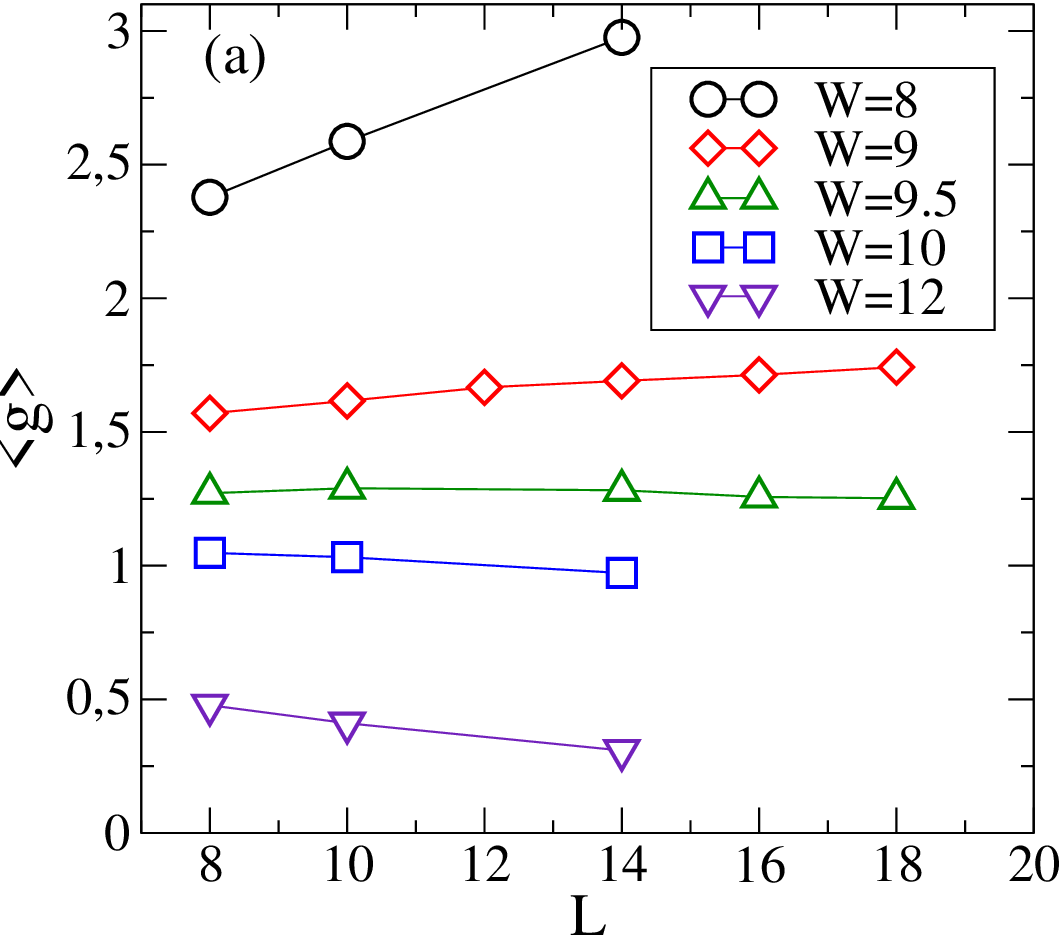}
\includegraphics[clip,width=0.35\textwidth]{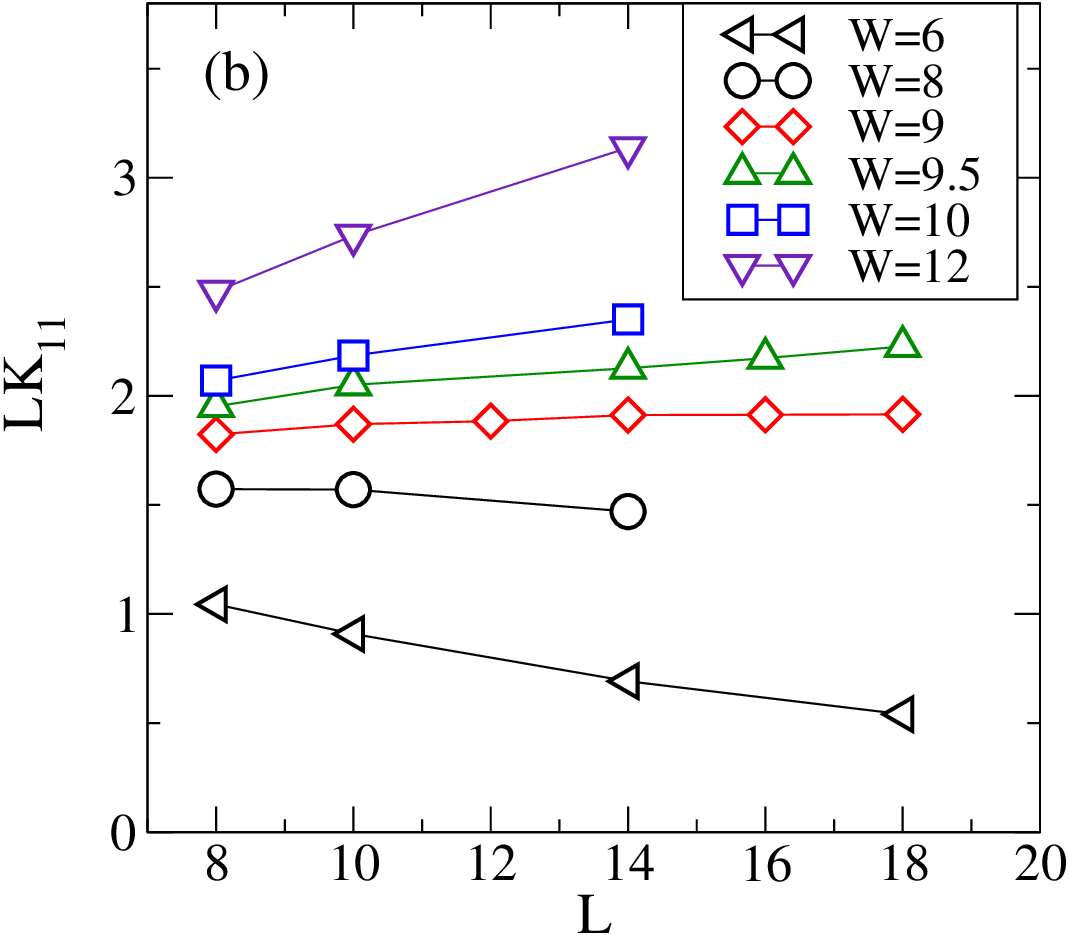}
\caption{Estimation of the critical point from the mean
conductance {$\langle g\rangle$} and from {$LK_{11}$}. Both
parameters $\langle g\rangle$ and $LK_{11}$ are $L$-independent at
$W=W_c$. $LK_{11}\propto L$ ($\propto 1/L)$ in the insulating
(metallic) regime, respectively. From data we conclude that
$9<W_c<9.5$ and use $W_c=9$ throughout the paper. }
\label{estimation}
\end{figure}

In the next step we analyze how the matrix $K$ changes when the
disorder increases. Typical results are given in
Fig.~\ref{W_dependence}. Our data show that the $L$ dependence of
the parameters $LK_{11}$ and $\gamma_{12}$ is different for
different transport regimes. While in the metallic regime
$LK_{11}$ decreases as $\sim 1/L$ and $\gamma_{12}$ converges to
unity when $L$ increases, qualitatively different behavior is
obtained at strong disorder. In the insulating regime $K_{11}$
converges to a non-zero $L$-independent constant when $L\to\infty$
(Fig.~\ref{3D_t04_k11}) and $K_{12}$ converges to zero  as
$K_{12}\sim 1/L$ (Fig.~\ref{3D_t04_k12}). This means that
$\gamma_{12}\sim 1/L$ in the insulating regime
(Fig.~\ref{loc_gamma}).

Figures 7 a,b  also show that there
exists a critical disorder $W=W_c$ where both $LK_{11}$ and
$\gamma_{12}$ are independent of $L$. Note that $\gamma_{12}<1$ at
$W=W_c$. We found that the critical value $\gamma_{12c}$ depends
on the anisotropy (Fig.~\ref{3D_t_cp}). For the present case
$t=0.4$, $\gamma_{12c}\approx 0.28$. The qualitative $L$
dependence in different transport regimes is summarized in Table
\ref{table:one}.

\begin{table}[b!]
\begin{center}
\begin{tabular}{lcccccc}
\hline
disorder      &  &&   $L\times {K_{11}}$    &&&   {$\gamma_{12}$}\\
\hline
$W\ll {W_c}$     &&   &        $\sim L^{-1}$ &  &&             1-${\cal{O}}(L^{-1})$   \\
${W_c}$          &  &&   {const}    & &&  {const} \\
$W\gg {W_c}$&&  & $\sim L$   &&       &     $\sim L^{-1}$\\
\hline
\end{tabular}
\end{center}
\caption{Typical $L$-dependence of the parameters $K_{11}$ and
$\gamma_{12}$ in the metallic, critical and localized regimes.}
\label{table:one}
\end{table}

We observe that the disorder dependence of  $K_{11}$ is consistent
with $K_{11}\propto 1/L^m$, where $m=2,1,0$ in the metallic,
critical and insulating limits, respectively, in agreement with
Ref.~[\onlinecite{chalker}]. Note that in contrast, $K_{11}\propto
1/L^2$ for all strengths of disorder in Q1D. This is a major
difference between Q1D and 3D. One can understand qualitatively
how the $L$ dependence of $K_{11}$ changes in the weak and strong
disorder limits on general grounds. If all channels are
equivalent, we expect the column matrix $v_{1a} \sim 1/\sqrt{N}$,
which satisfies the unitary condition $\sum_{a=1}^N |v_{1a}|^2
=1$. This leads to $K_{11}\sim 1/N=1/L^2$ in the metallic limit.
On the other hand, if the localization length $\xi \sim 1$, then
on any cross-section at a given $L_z$, we expect only a few sites
on the back side of the sample to be `illuminated' by an incoming
wave, so we expect $v_{1a}\sim \delta_{1a}$. This leads to
$K_{11}\sim 1$, independent of $L$. Similarly, since all
$K_{1a}\propto 1/L^2$ in the metallic regime, we expect
$\gamma_{12}\sim 1$ in the metallic regime. However, in the
insulating regime, we have not found a simple physical argument
why $K_{12}\propto 1/L$ and hence $\gamma_{12}\propto 1/L$. We
also find numerically in the insulating regime that for $1\ll \xi
\ll L$,  $K_{11}\sim 1/\xi$. The structure of the eigenvector
$v_{1a}$ that gives rise to $K_{11}\sim 1/\xi$ and $K_{12}\sim
1/L$ in the region $1\ll \xi \ll L$ is highly non-trivial, and
deserves further study.

Figure \ref{3D_t04_k11} confirms our claim that $K_{11}>0$ in the
insulating regime. Its limiting value, $\tK=\lim_{L\to\infty}
K_{11}(L)$ can be used as an order parameter for the scaling
analysis of the Anderson transition. It is evident that $\tK=0$
for $W<W_c$ and $\tK>0$ for $W>W_c$. We show in
Fig.~\ref{estimation} the $L$-dependence of $LK_{11}$ and of the
mean conductance $\langle g\rangle$ for various strengths of
disorder. Similar behavior is shown in Fig.~\ref{loc_gamma} for
the parameter $\gamma_{12}$. One sees that all three parameters,
$\langle g\rangle$, $LK_{11}$ and $\gamma_{12}$,  could be used
for the estimation of the critical disorder $W_c$, at which  none
of them depends on the system size.

\begin{figure}[t]
\includegraphics[clip,width=0.35\textwidth]{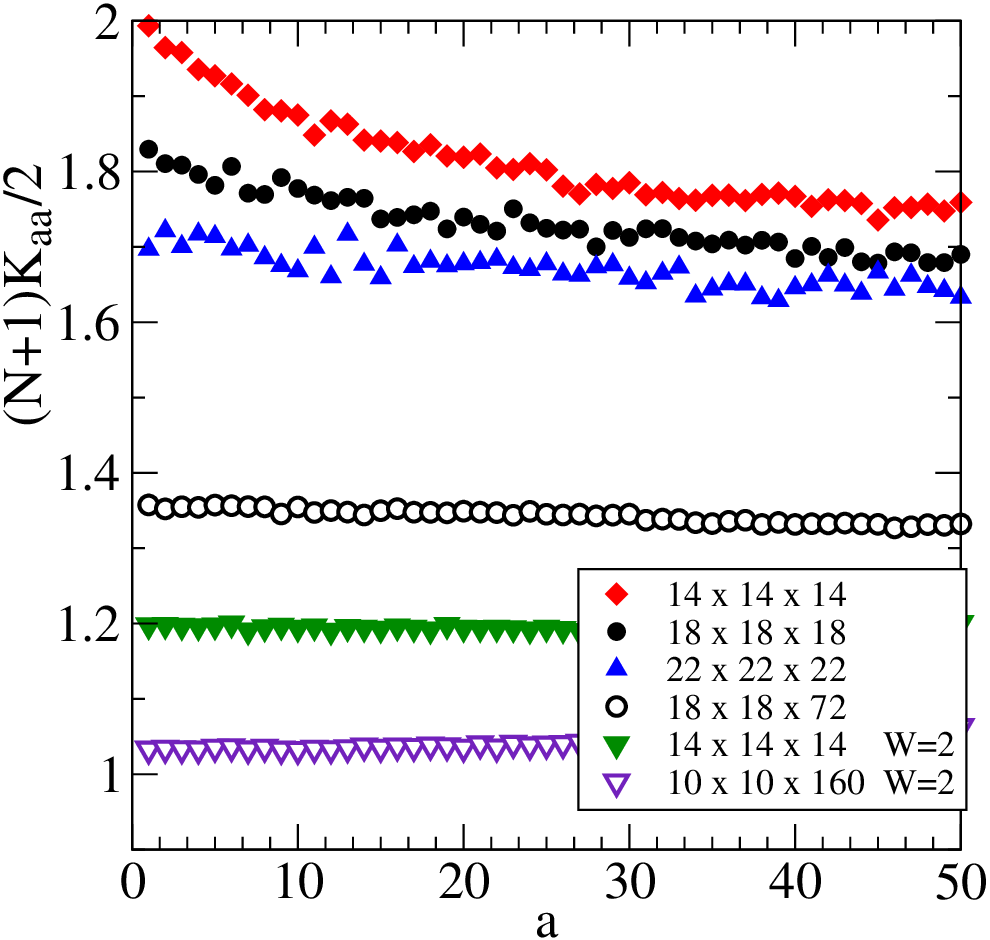}
\caption{$a$-dependence of $K_{aa}\times(N+1)/2$ for 3D and Q1D
systems with $W=2$, (mean free path $l\approx 9.2$) and for $W=4$
($l\approx 1.8$).  $K_{aa}$ are larger than is predicted by the
DMPK theory. Good agreement with DMPK is observed only for Q1D
systems with very small disorder ($W=2$). } \label{metal_kaa}
\end{figure}

\begin{figure}
\includegraphics[clip,width=0.35\textwidth]{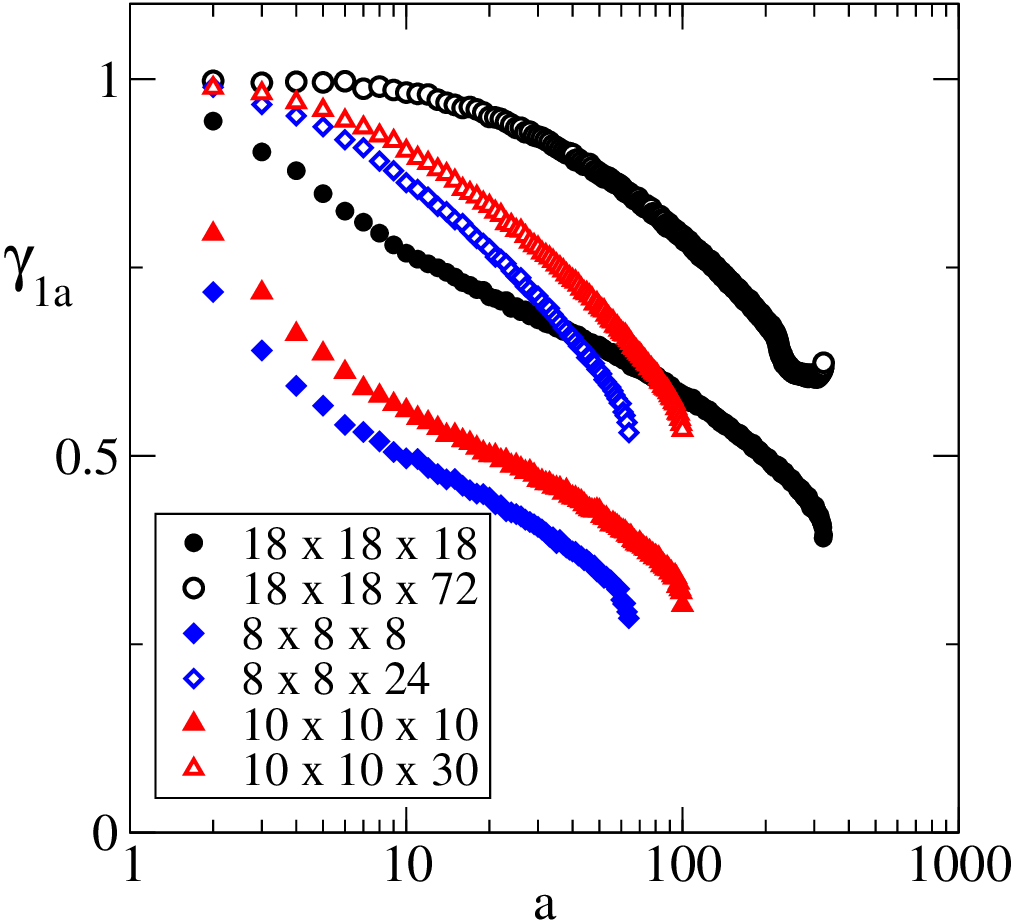}
\caption{$a$-dependence of $\gamma_{1a}$ for the  metallic regime
($W=4$). Finite size effects are clearly visible so that
$\gamma_{12}=1$ only in the limit of large system size. As
expected, convergence is better for Q1D systems than for cubes.
Nevertheless, data confirm that $\gamma_{1a}$ converge to 1 in the
limit $L\to\infty$ even when $K_{11}$ and $K_{12}$ do not converge
to values assumed by DMPK given in Eq.~(\ref{dmpk-q1d}).}
\label{metal_gamma1a}
\end{figure}

\subsection{Q3: Index dependence of $K_{ab}$}

Again we begin with weak disorder. To compare 3D and Q1D  systems,
we show in Fig.~\ref{metal_kaa} the parameters $K_{aa}$ as a
function of the index $a$. It is clear that the 3D data differ
considerably from the DMPK value $2/(N+1)$. In
Fig.~\ref{metal_gamma1a} we show the ratio
$\gamma_{1a}=2K_{1a}/K_{11}$ for various $L$ and compare it with
Q1D numerical data. It is evident that $\gamma_{1a}$ converges to
1 when the system size increases, in spite of the fact that both
$K_{1a}$ and $K_{11}$ differ from the DMPK values of
Eq.~\ref{dmpk-q1d}. The convergence is much slower in 3D than in
Q1D systems. Also, $\gamma_{1a}$ converges slower for larger $a$.
We conclude that although 3D metals are qualitatively similar to
Q1D metals, there are quantitative differences that need to be
explored further.

\begin{figure}
\includegraphics[clip,width=0.35\textwidth]{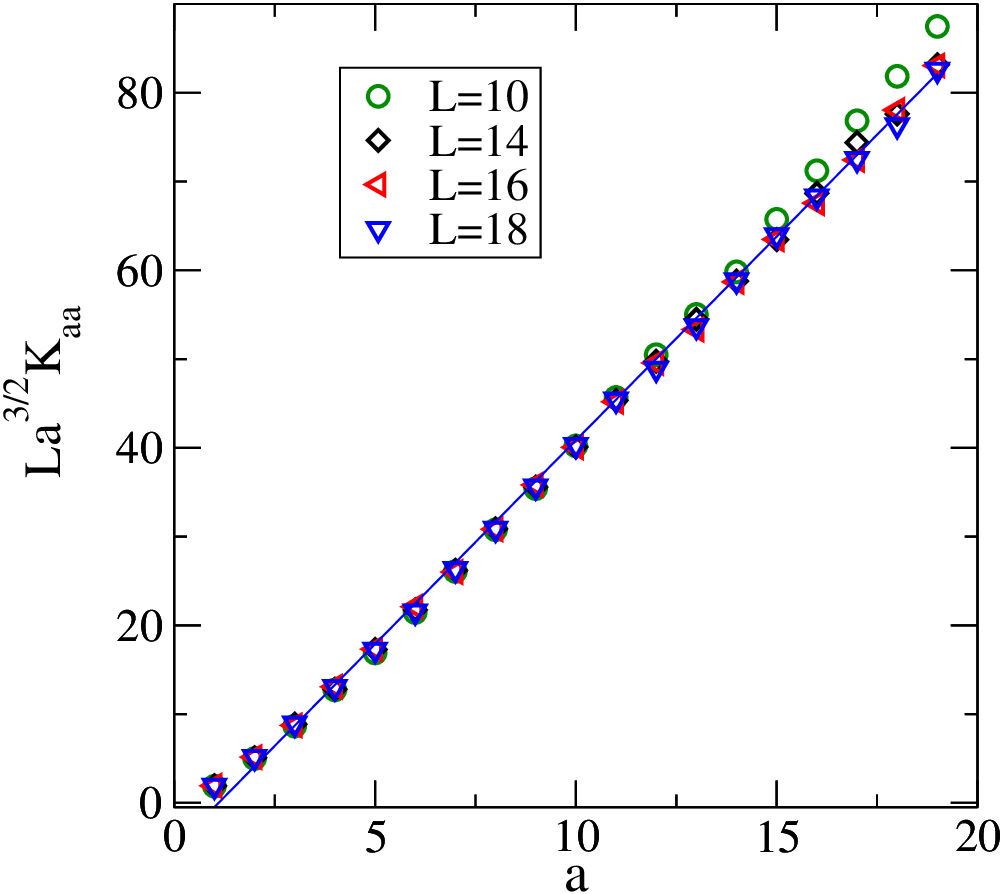}
\caption{Critical point: We plot $La^{3/2}K_{aa}$ to show that
$LK_{aa}\propto a^{-1/2}$ are $L$-independent for all $a\le 20$}
\label{k_cp_a}
\end{figure}

\begin{figure}
\includegraphics[clip,width=0.35\textwidth]{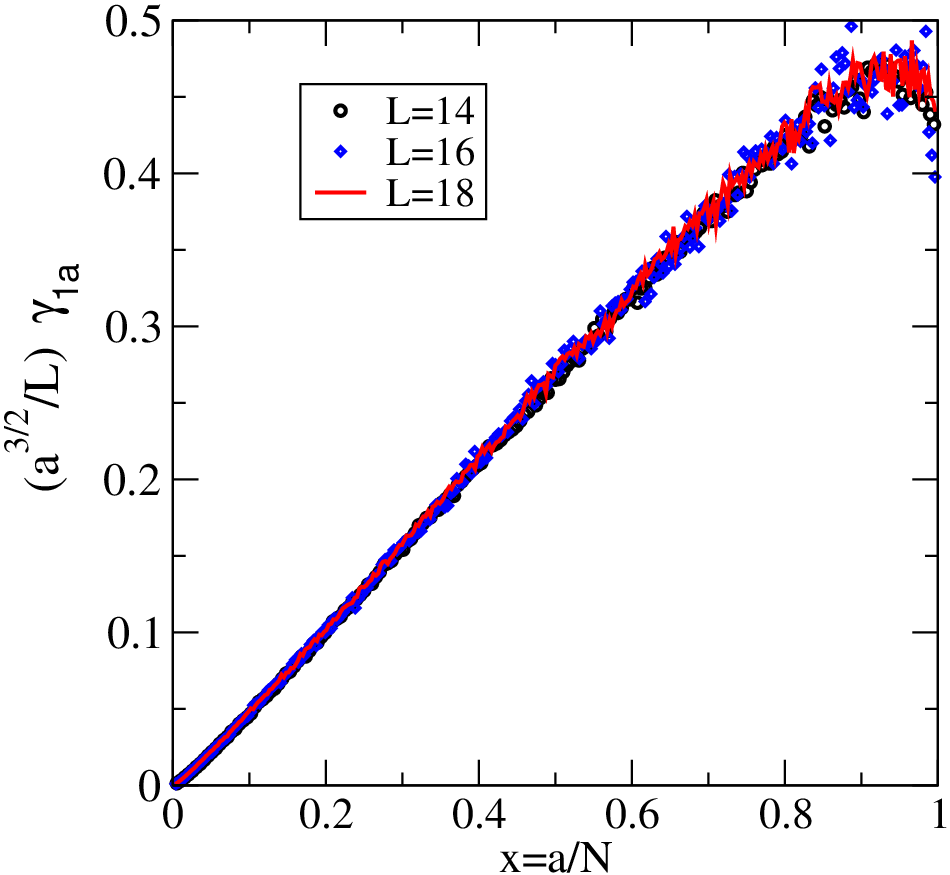}
\caption{Critical point. From the linear dependence
$(a^{3/2}/L)\gamma_{1a}\propto a/N$ we conclude that
$\gamma_{1a}\propto 1/a^{1/2}L$. ($N=L^2$)}
\label{k_cp_b}
\end{figure}

In the critical regime, Figures \ref{k_cp_a} and \ref{k_cp_b} show
that the $a$ and $b$ dependence of the matrix elements $K_{ab}$
can be described by simple functions: $K_{aa}\sim K_{11}/a^{1/2}$,
$\gamma_{1a}\sim 1/La^{1/2}$. Although we did not analyze all
matrix elements in detail, we believe that the data presented here
support our expectation that all matrix elements $K_{ab}$ can be
expressed in terms of $K_{11}$, $K_{12}$ and some simple function
of the indices $a$ and $b$.

%\begin{figure}
%\includegraphics[clip,width=0.5\textwidth]{g_ii1.eps}
%\caption{%
%${\gamma_{a,a+1}\approx a^{1/4}}$. Data show that
%$\gamma_{ab}$ does not depend on the system size $L$ at the critical point.}
%\label{k_cp_c}
%\end{figure}

\begin{figure}[t]
\includegraphics[clip,width=0.45\textwidth]{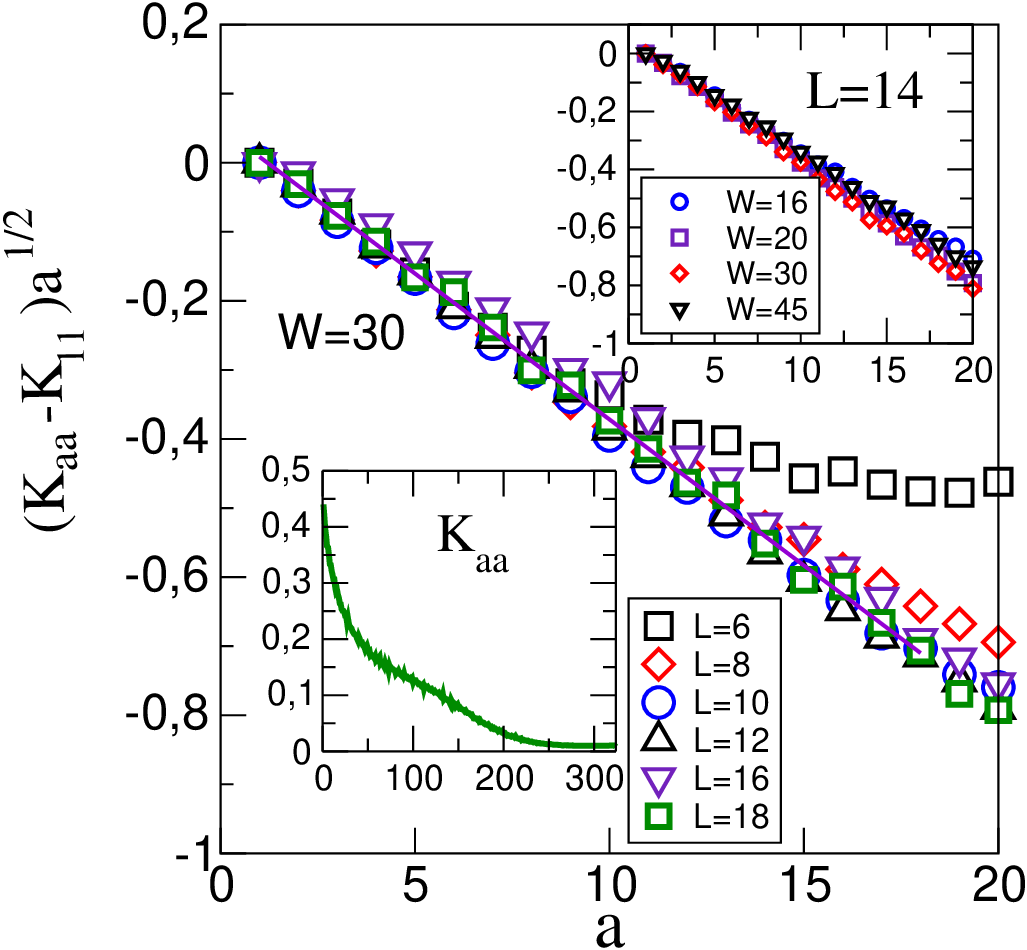}
\caption{Insulating regime: $a$-dependence of $K_{aa}$ for 3D
systems with $W=30$. To see the $a$-dependence more clearly, we
plot the difference $[K_{aa}-K_{11}]\sqrt{a}$ which behaves as
$\sim -a$ for small $a$. Solid line is a linear fit, from which we
have  $K_{aa}=K_{11}+0.05-0.04/\sqrt{a}$ for $L=18$ and $a\le 18$
($K_{11}=0.44$).
Right inset presents data for $L=14$ and various strength of the disorder.
Left inset shows $K_{aa}$ for $L=18$.
} \label{loc_b}
\end{figure}

\begin{figure}[t]
\includegraphics[clip,width=0.45\textwidth]{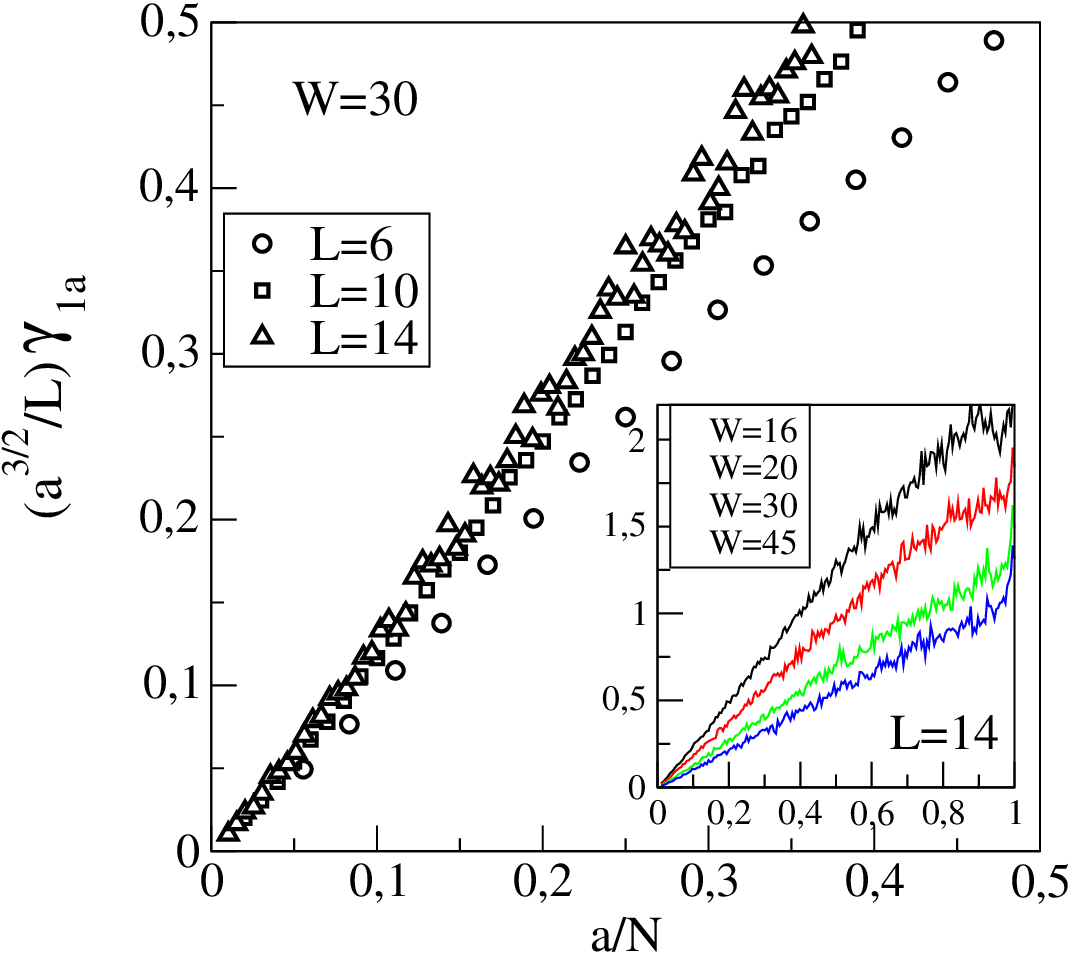}
\caption{Insulating regime ($W=30$): $a$-dependence of
$\gamma_{1a}$ ($L=6$, 10 and 14). Similar to the critical point,
data shows that  $a^{3/2}/L\times \gamma_{12}\sim a/N$ so that
$\gamma_{1a}\sim 1/(\sqrt{a}L)$.
Inset shows data for $L=14$ and various strength of the disorder.
} \label{loc_c}
\end{figure}

In the insulating regime  we find a simple $a$-dependence of the
\textsl{difference} \be K_{aa}-K_{11}\approx  -c\sqrt{a} \ee with
$|c|\sim 0.04$ (Fig.~\ref{loc_b}). We also found,
(Fig.~\ref{loc_c}), that for $a<N/4$, $\gamma_{1a}\sim
1/\sqrt{a}L$, very close to the value obtained at the critical
point \cite{note}.

There is an interesting correspondence between the $a$-dependence
of $K_{aa}$ and the  $a$-dependence of the parameters $x_a$
defined through the parameters \cite{pichard-nato} \be
\lambda_a=\sinh^2 x_a \ee which is summarized in Table
\ref{table:two}.

\begin{table}
\begin{tabular}{llll}
\hline
\hline
~~   &metal~~~~         &    critical point~~~   &   insulator~\\
\hline
$x_a$~~~~~~~~~  &   $a x_1$   &    $\sqrt{a} x_1$     &   $x_1+c\sqrt{a}$\\
$K_{aa}$ &  $K_{11}$   &    $K_{11}/\sqrt{a}$   &   $K_{11} -c/\sqrt{a}$\\
\hline
\hline
\end{tabular}
\caption{Index dependence of $K_{aa}$ obtained in the present work
compared to  $x_a$ \cite{JPCM} in the metallic, critical and
insulating regimes.} \label{table:two}
\end{table}

We find that the index dependence of $K_{aa}$ can be ignored in
the metallic regime. The $a$-dependence is more pronounced at the
critical point and in the localized regime.  On the other hand,
higher channels ($a\gg 1$) do not contribute to the transport
either at the critical point or in the localized regime, so the
actual values of $K_{ab}$ for large $a$ and $b$ are not important.
Therefore, we conclude that the weak index dependence of the
matrix elements $K_{ab}$ is less relevant for transport properties
compared to the dependence of the matrix elements on disorder in
the $L\rightarrow \infty$ limit.

It is also worth mentioning  that since we are interested only in
$L$-independent quantities at the critical point, the
$a$-dependence of any parameter is relevant  only  for $a\le L$.
When $a$ becomes comparable to $L$, we can not distinguish the
true $a$-dependence from finite size effects.

\subsection{Q4: Simple model for K}

Finally we ask the question if it is possible to construct a
simple model of $K_{ab}$ with only a small number of independent
parameters. We just concluded in the previous section that the
weak index dependence of the matrix at strong disorder is not very
important. We expect that the crude approximations
\begin{equation} \label{crude}
K_{aa}\approx K_{11}\;\;\;  \textrm{and} \;\;\; \gamma_{ab}\approx
\gamma_{12}
\end{equation}
capture the major qualitative features of the matrix
$K$.

\begin{figure}
\includegraphics[clip,width=0.35\textwidth]{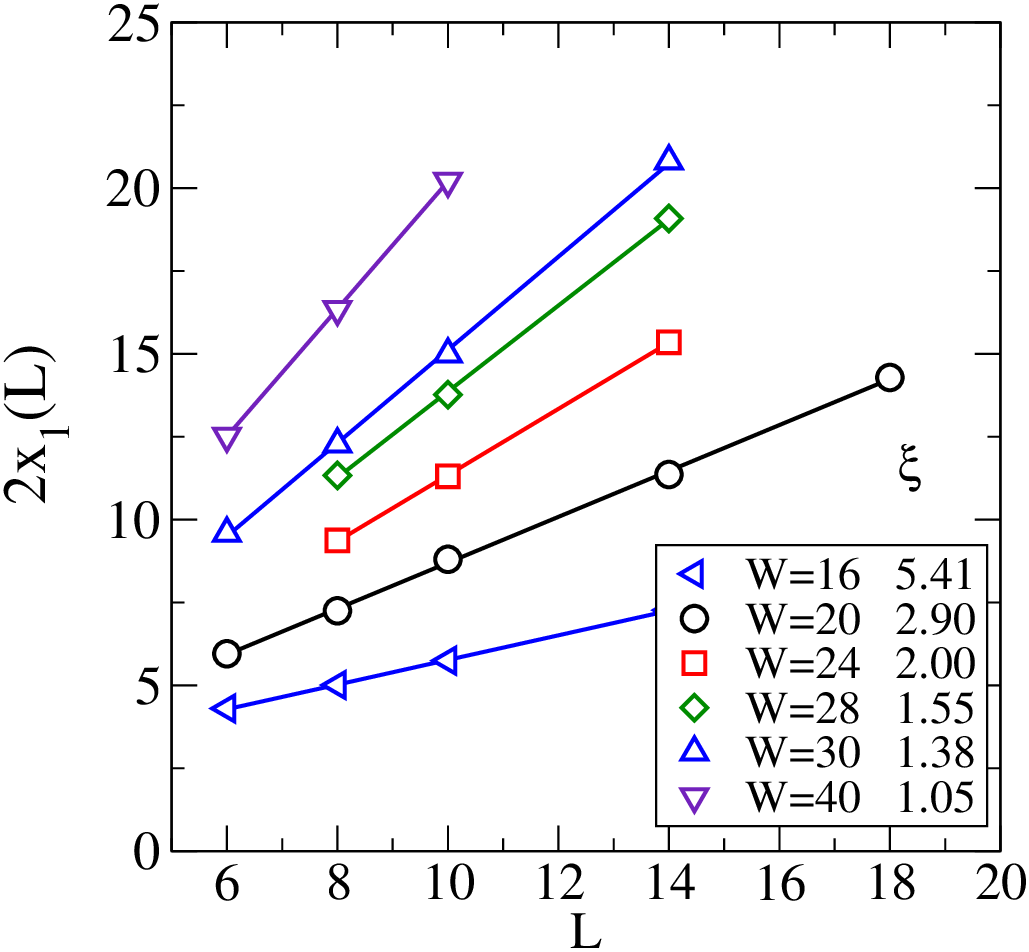}
\caption{%
Insulating regime: Estimation of the localized length $\xi(W)$
from the linear $L$ dependence $x_1(L)=\textrm{const}+L/\xi$.
($\lambda_1=\sinh^2 x_1$). Values of $\xi$ are given in the
legend. $g\approx \cosh^{-2}x_1\approx \exp -2L/\xi$. }
\label{3D_z1}
\end{figure}

\begin{figure}
\includegraphics[clip,width=0.35\textwidth]{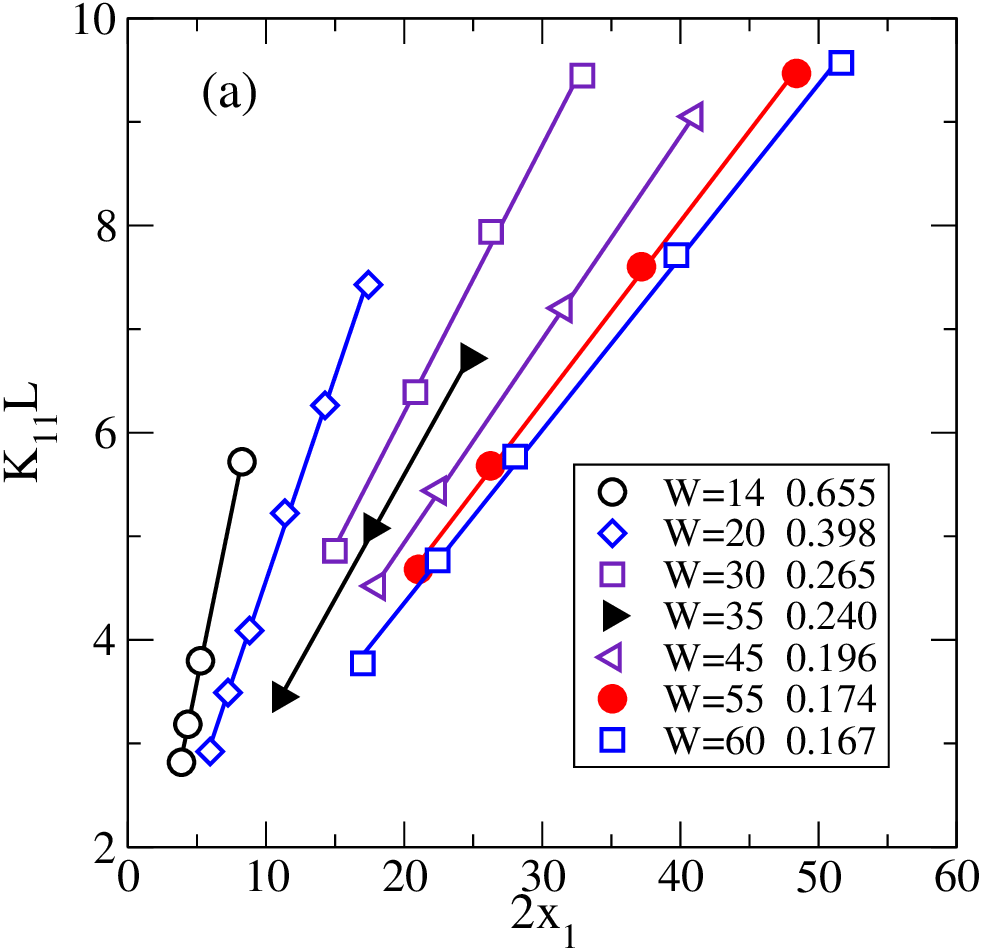}
\includegraphics[clip,width=0.35\textwidth]{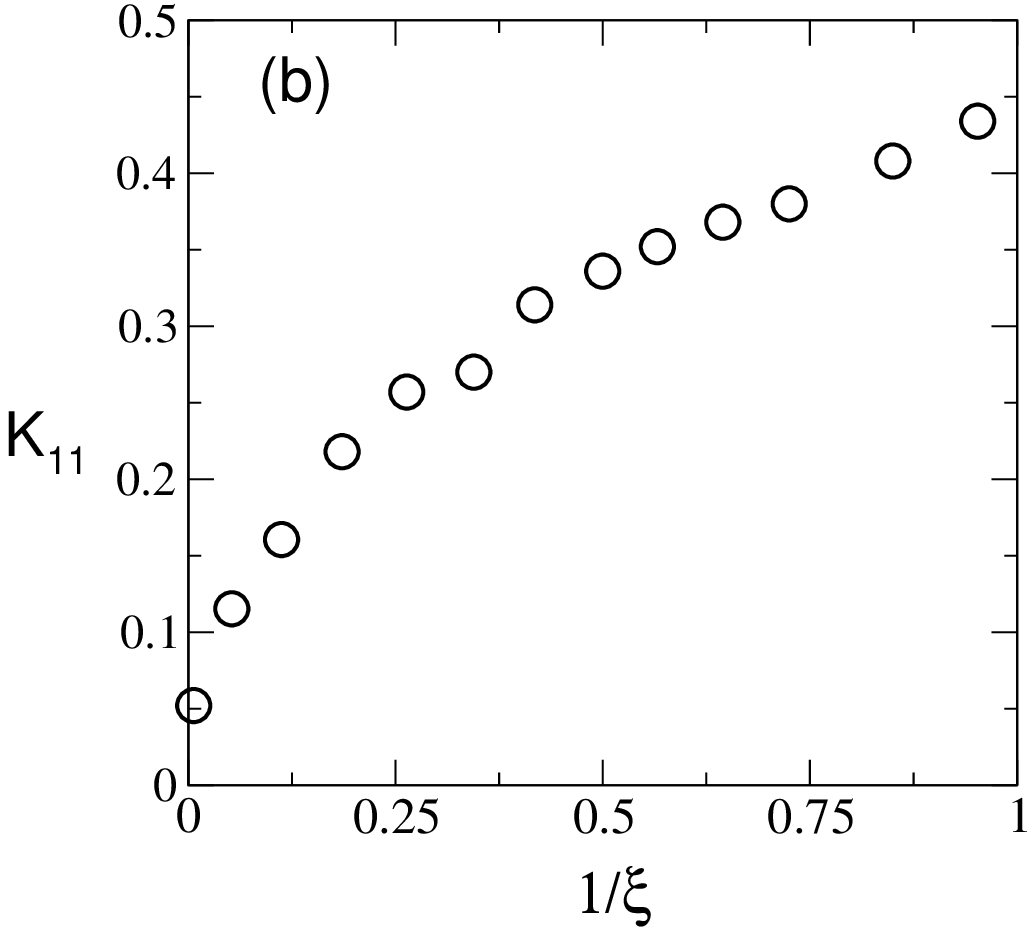}
\caption{%
Figure (a) shows how $LK_{11}$ depends on  $x_1$ for various
disorder in the localized regime. Data confirm linear relation $LK_{11}\propto x_1$.
Figure (b) shows how limiting values of $K_{11}$,
obtained from the $L$-dependence of $K_{11}(L)$
(Fig.~\ref{3D_t04_k11}), depend on values of $1/\xi$
(Fig.~\ref{3D_z1}). Data confirms that there is an unambiguous
$\xi$-dependence of $K_{11}$.
} \label{3D_univ}
\end{figure}

Approximation (\ref{crude}) introduces two new parameters,
$K_{11}$ and $\gamma_{12}$.
We show that both  parameters
$K_{11}$ and $\gamma_{12}$ are unambiguous functions of the
localization length. To do so, we calculated the limiting values
of $K_{11}$ (Fig.~\ref{3D_t04_k11}) and of the parameter $x_1$
(Fig.~\ref{3D_z1}) and plot  $K_{11}$ \textsl{versus} $\xi$
(Fig.~\ref{3D_univ}). We observe  that the $\xi$ dependence of
$K_{11}$ depends also on the anisotropy parameter $t$ (data not
shown).

\begin{figure}
\includegraphics[clip,width=0.35\textwidth]{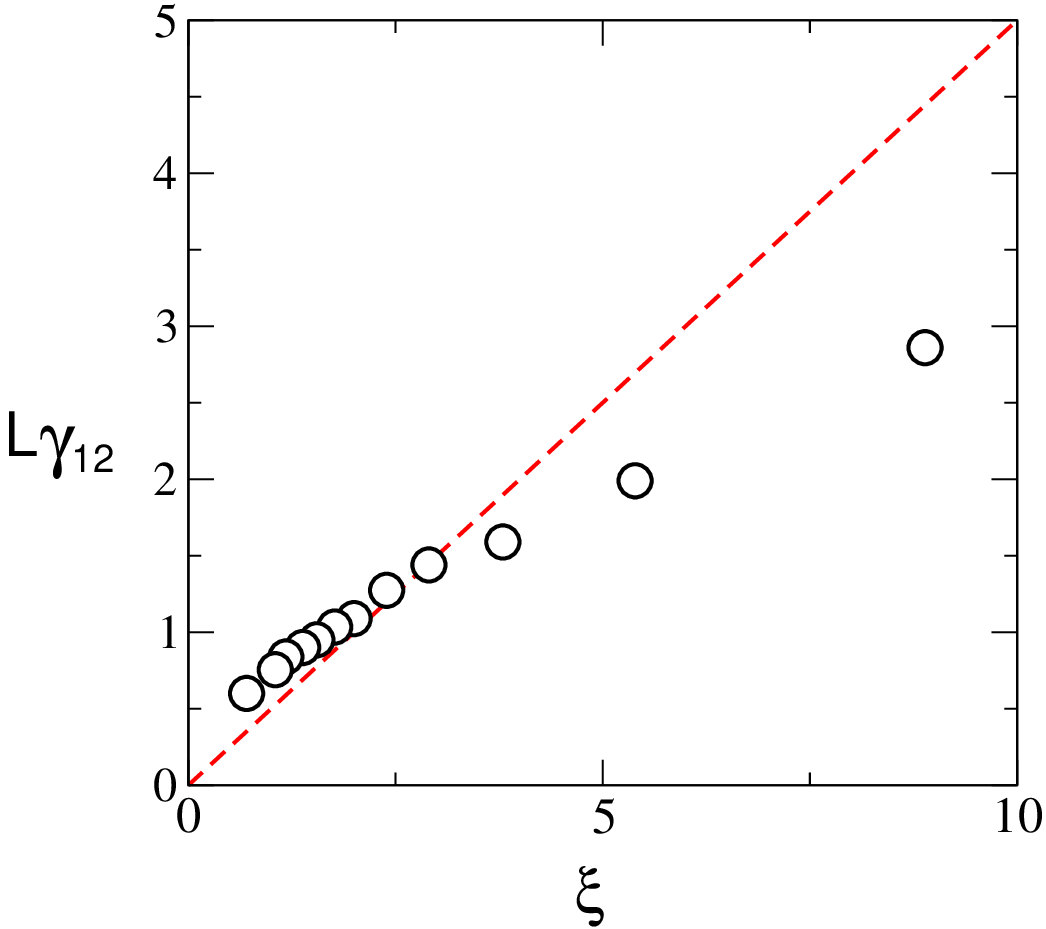}
\caption{%
$L\gamma_{12}$ defined by Eq.~(\ref{Lg}) as a function of $\xi$.
Data confirms  that there is an unambiguous $\xi$-dependence of
$\gamma_{12}$. This is important for the formulation of the single
parameter scaling theory. Dashed line is the linear dependence
$\gamma_{12}=\xi/2L$, considered in [\onlinecite{mmwk}]. }
\label{3D_univ_Lg}
\end{figure}

In the same way, we analyzed parameter $\gamma_{12}$. As was shown
in Fig.~\ref{loc_gamma}, we expect $\gamma_{12}\sim L^{-1}$ in the
localized regime. Data in Figures \ref{3D_t04_k11} and
\ref{3D_t04_k12} support this assumption. To estimate the disorder
dependence of  $\gamma_{12}$, we plot in Fig.~\ref{3D_univ_Lg} the
quantity \be\label{Lg} \lim_{L\to\infty}
L\gamma_{12}=2L~\displaystyle{\frac{\lim_{L\to\infty}
K_{12}}{\lim_{L\to\infty} K_{11}}} \ee which indeed increases
linearly with localization length in the strongly localized
regime.

\medskip

%Fig. \ref{3D_t04_Gamma} shows also how the ratio $\Gamma/\gamma_{12}$ depends on the disorder for 3D insulating regime. The typical values $\Gamma/\gamma_{12}$ are smaller than 4, assumed in Ref. \cite{mmwk}.  Data show that $\Gamma/\gamma_{12}$  is unambiguous function of $W$ which again supports single parameter scaling hypothesis.

\begin{figure}
\includegraphics[clip,width=0.35\textwidth]{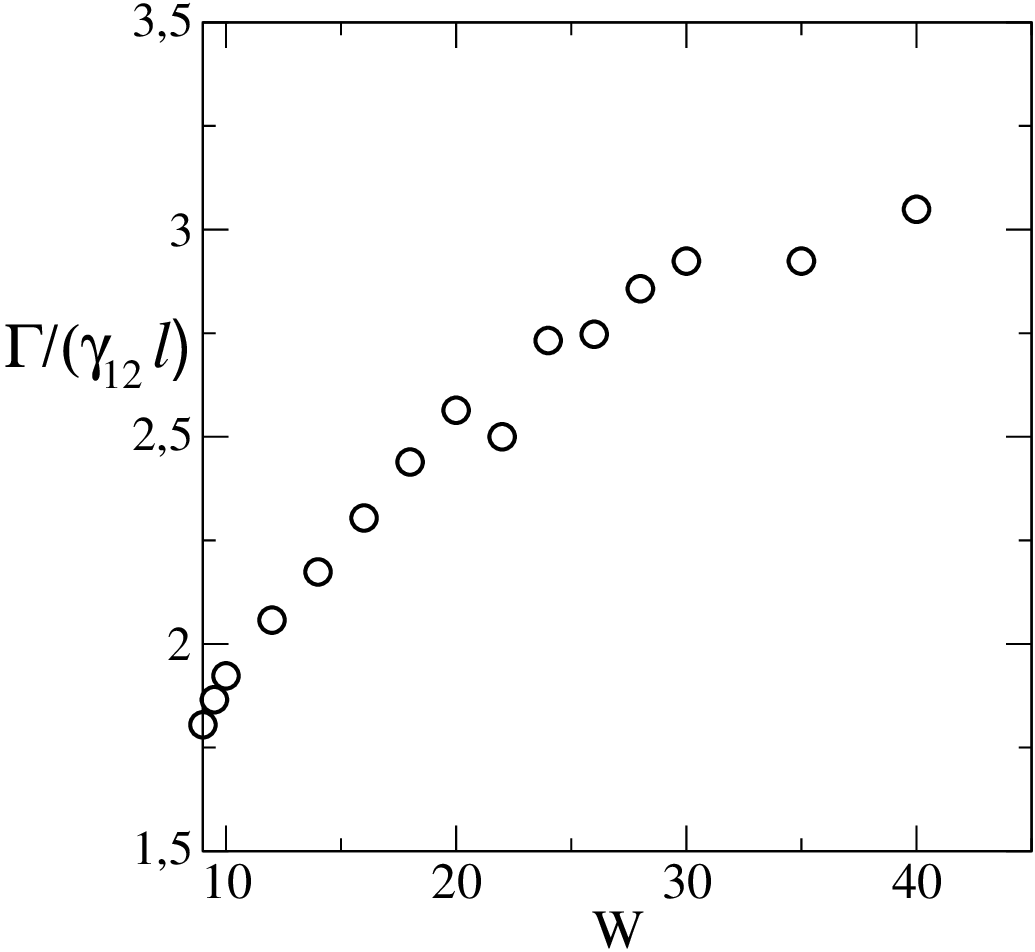}
\caption{%
Ratio $\Gamma/(\gamma_{12}\ell)=1/2LK_{12}$, given by Eq. \ref{Gamma}
for a cubic system, as a function of disorder for 3D systems in
the insulating regime $W>W_c$. } \label{3D_t04_Gamma}
\end{figure}

\section{Solution of the generalized DMPK equation in the strong disorder limit}

While modeling the full $\gamma_{ab}$ at the critical point needs
more careful numerical studies, the insulating limit is simpler
and provides a test case for the generalized DMPK equation (8). It
predicts that the logarithmic interaction between the transmission
eigenvalues $\lambda_a$ vanishes as $1/L$ in the insulating limit.
In the rest of the paper we will test this prediction by
evaluating the full distribution of conductances in the insulating
limit for a 3D conductor as described by Eq.~(8), using the simple
approximate model for $K$ suggested by our numerical studies,
namely
\begin{equation}\label{simple}
%K_{aa}\approx K_{11}\sim 1/\xi~~{\rm and}~~ \gamma_{ab}\approx
K_{aa}\approx K_{11}~~{\rm and}~~ \gamma_{ab}\approx
\gamma_{12}\sim \textrm{const.}\frac{\xi}{L}.
\end{equation}
It is useful to introduce  another parameter, \be
\Gamma=\frac{\ell}{L_zK_{11}} \ee and the ratio \be\label{Gamma}
\frac{\Gamma}{\gamma_{12}}=\frac{\ell}{L_zK_{11}\gamma_{12}}
=\frac{L}{L_z}~\frac{\ell}{2LK_{12}}.
\ee
$\Gamma\sim 1/L_z \ll 1$ in the insulating regime, both in
3D ($L=L_z$) and Q1D ($L_z\gg L$) systems. It measures the
strength of the disorder. The ratio $\Gamma/\gamma_{12}$ reduces
to $\Gamma$ in the Q1D limit when $\gamma_{12}=1$. For 3D
($L_z=L$) we see in Fig.~\ref{3D_t04_Gamma} that
$\Gamma/\gamma_{12}\sim L\ell/L_z$  varies smoothly between about $2\ell$
and $3.5\ell$ when disorder increases from $W_c$ to infinity. Thus, the
strength of disorder is characterized predominantly by the parameter
$\Gamma$. For definiteness, we will use $\Gamma/\gamma_{12}=2$
appropriate for a cubic system when needed for comparison with
numerical results \cite{noteG/g}.

Note that fluctuations in $k_{aa}$, ignored in the model, would
lead to fluctuations of $\Gamma$ and $\gamma_{12}$ as well, but
will not change either the length or the disorder dependence of
these parameters.

A brief description of the results has appeared in
[\onlinecite{mmwk}]. Here we provide many of the details.

We rewrite the generalized DMPK as \cite{mu-go-02}
\begin{equation}\label{genDMPK}
\frac{\partial p_s(x)}{\partial (L_z/\ml)} =\frac{1}{4}\sum_a K_{aa}
\frac{\partial}{\partial x _a}\left[\frac{\partial p}{\partial
x_a}+p\frac{\partial \Omega}{\partial x_a}\right]
\end{equation}
where 
\begin{equation}
\Omega\equiv -\sum_{a<b}^N \ln |\sinh^2 x_b-\sinh^2
x_a|^{\gamma_{ab}}-\sum_{a=1}^N \ln \sinh 2x_a.
\end{equation}
We define $P=e^{-\Omega/2}\Psi$.  Then following
[\onlinecite{beenakker}], $\Psi$ satisfies the imaginary time
Schrodinger equation $-\partial\Psi/\partial
(L_z/\ml) =(\mathcal{H}-U)\Psi$ with
\begin{equation}\label{ch}
\begin{array}{ll}
\mathcal{H}=&\displaystyle{-\frac{K_{11}}{4}\sum_a
\left[\frac{\partial^2}{\partial x^2_a}+ \frac{1}{\sinh^2
2x_a}\right]}\\
~&\displaystyle{+ \eta\sum_{a <
b}\left[\frac{1}{\sinh^2(x_a-x_b)}+\frac{1}{\sinh^2(x_a+x_b)}\right]},
\end{array}
\end{equation}
and $U$ constant, where the strength of the interaction is given
by
\begin{equation}
\eta=\frac{K_{11}}{4}\gamma_{12}(\gamma_{12}-2).
\end{equation}

The interaction term in Q1D vanishes for the unitary case, when
$\gamma^{Q1D}_{12}=\beta=2$. Note that the interaction also
vanishes in the limit $\gamma_{12}\rightarrow 0$. In the 3D
insulating case, $\gamma_{12}\sim \xi/2L \ll 1$, and the
interaction can be considered negligible, for all symmetries. We
can therefore use the $\gamma_{12}=2$ solution of
[\onlinecite{beenakker}] for our 3D insulators. The solution in
the insulating limit is then given by \cite{beenakker} $P=e^{-H}$,
with
\begin{equation}\label{ham}
\begin{array}{ll}
H=&-\displaystyle{\sum_{a>b}^N\left[\frac{1}{2} \ln |\sinh^2 x_b-\sinh^2
x_a|^{\gamma_{12}}+\ln |x^2_b- x^2_a|\right]}\\
~&\displaystyle{-\sum_{a=1}^N \left[\frac{1}{2}\ln \sinh 2x_a+\ln
x_a-\Gamma x^2_a\right]}.
\end{array}
\end{equation}

The replacement of $\beta=2$ in Ref.~[\onlinecite{beenakker}] by
$\gamma_{12}\rightarrow 0$ in Eq.~(\ref{ham}) has the consequence
that while all $\langle x_a\rangle\gg 1$ in the insulating regime,
the difference $s=\langle x_{a+1}-x_a\rangle$ is \textsl{not} of
the same order as $\langle x_a\rangle$. For example, if we keep
only the first two levels, the saddle-point solutions for $x_1$
and $x_2$ give $\langle x_1\rangle\sim L_z/\xi$ and $\langle
x_2-x_1\rangle\ll \langle x_1\rangle$. We therefore do not assume
that $x_2\gg x_1$. However, we do make the simplifying
approximation that $ \ln |\sinh^2 x_a-\sinh^2 x_b|\approx \ln
\sinh^2 x_a$ and $\ln |x^2_a- x^2_b|\approx \ln x^2_a$ for $a>2$
and $a>b$. Eq.~(\ref{ham}) then becomes
\begin{equation}\label{happ}
H\approx H_1+\sum_{a=2}^N\left[V(x_a)
-\gamma_{12}(a-2)f(x_a)\right],
\end{equation}
where
\begin{equation}
H_1  =  -\ln|x^2_2-x^2_1| + \Gamma x^2_1-\frac{1}{2}\ln \sinh 2
x_1 -\ln x_1,
\end{equation}
\begin{equation}\label{V}
V(x) = \Gamma x^2-\frac{1}{2}\ln \sinh 2 x -\ln x - \gamma_{12}\ln
\sinh x,
\end{equation}
\begin{equation}\label{f}
f(x) = \ln \sinh x+k\ln x; \;\;\; k=\frac{2}{\gamma_{12}}.
\end{equation}

We can now use the method developed in
Refs.~[\onlinecite{mu-wo-99,mu-wo-go-03}] to obtain the full
distribution $P(g)$.

\section{$P(g)$ in 3D in the insulating limit}

As in [\onlinecite{mu-wo-99,mu-wo-go-03}], we separate out the
lowest level $x_1$ and treat the rest as a continuum beginning at
a point $x_2$. Then
\begin{equation}
H=H_1+\int_{x_2}^{b}dx'\sigma (x')\left[V(x')-\gamma_{12}
f(x')\int_{x_2}^{x'}dx''\sigma (x'')\right]
\end{equation}
where we have used
\begin{equation}
(a-2)=\int_{x_2}^x dx'\sigma (x').
\end{equation}
The density satisfies the normalization condition
\begin{equation}
\int_{x_2}^{b}\sigma (x)dx=N-1.
\end{equation}
$P(g)$ can be obtained from $H(\{x_a\})$ as
\begin{equation}
P(g)=\int\cdots\int\prod_a dx_a e^{-H}\delta (g-\sum_a {\rm
sech}^2 x_a)
\end{equation}
where the $\delta$-function represents the Landauer formula for
conductance. It turns out that because of the nonlinear dependence
of the Hamiltonian on the density, the complex delta function
representation used in [\onlinecite{mu-wo-99,mu-wo-go-03}] is not
suitable for the present case. We therefore use a real
representation
\begin{equation}
\delta (x-a)=\lim_{\lambda\rightarrow
0}\frac{1}{\lambda\sqrt{\pi}} e^{-(x-a)^2/\lambda^2}.
\end{equation}
Following [\onlinecite{mu-wo-99,mu-wo-go-03}], $P(g)$ may be
expressed as
\begin{equation}
P(g) =\int D[\sigma(x)]\int_0^{\infty}dx \int_{x_1}^{\infty} dx_2
e^{-F(x_1,x_2;\sigma(x))}
\end{equation}
where the free energy functional $F$ is given by
\begin{equation}
\begin{array}{ll}
F=&\displaystyle{H_1+\int_{x_2}^{b}dx'\sigma (x')\left[V(x')-\gamma_{12}
f(x')\int_{x_2}^{x'}dx''\sigma (x'')\right]}\\
~&\displaystyle{+
\frac{1}{\lambda^2}\left[(g-h_1)-\frac{\kappa}{2}\right]^2}
\end{array}
\end{equation}
where
\begin{equation}
\kappa\equiv 2\int_{x_2}^{b} dx'\sigma (x')h(x')
\end{equation}
and
\begin{equation}
h(x)\equiv {\rm sech}^2 x; \;\;\; h_1\equiv {\rm sech}^2 x_1.
\end{equation}
The saddle point density is to be obtained by minimizing $F$ with
respect to $\sigma(x)$, subject to the normalization condition. We
therefore define
\begin{equation}
\mathcal{F}\equiv F-\Lambda\left[\int_{x_2}^{b}\sigma (x)
dx-(N-1)\right]
\end{equation}
and minimize $\mathcal{F}$. It is useful to rewrite the free
energy using the normalization condition in a way that removes the
upper limit from the resulting equation. We use
\begin{eqnarray}
&&\frac{\delta}{\delta\sigma} \int_{x_2}^{b}dx'\sigma
(x')f(x')\int_{x_2}^{x'}dx''\sigma (x'')\nonumber \\ & = &
\frac{\delta}{\delta\sigma}\int_{x_2}^{b}dx'\sigma
(x')f(x')\left[(N-1)-\int_{x'}^{b}dx''\sigma (x'')\right]
\nonumber \\ & = & f(x) \int_{x_2}^{x}dx'\sigma
(x')-\int_{x_2}^{x}dx'\sigma (x')f(x')
\end{eqnarray}
to obtain
\begin{equation}\label{YL}
Y(x)-\Lambda = \gamma_{12} \left[f(x)\int_{x_2}^{x}dx'\sigma
(x')-\int_{x_2}^{x}dx'\sigma (x')f(x')\right]
\end{equation}
where we have defined
\begin{equation}
Y(x)\equiv V(x)-
\frac{2}{\lambda^2}\left[(g-h_1)-\frac{\kappa}{2}\right]h(x).
\end{equation}
Eq.~(\ref{YL}) evaluated at $x=x_2$ fixes $\Lambda = Y(x_2)$.
Taking a derivative of Eq.~(\ref{YL}) with respect to $x$
(represented by a prime) gives
\begin{equation}\label{Y'}
Y'(x)= \gamma_{12} f'(x)\int_{x_2}^{x}\sigma (y)dy
\end{equation}
Evaluated at $x=x_2$, this fixes $x_2$ as the beginning of the
continuum
\begin{equation}
Y'(x_2)= 0.
\end{equation}
Taking another derivative with respect to $x$, it is now possible
to obtain the density
\begin{equation}\label{sigma}
\sigma(x)=\frac{1}{\gamma_{12}}\left(\frac{Y'(x)}{f'(x)}\right)'.
\end{equation}
We check that plugging $\sigma(x)$ in Eq.~(\ref{sigma}) back to
Eqs.~(\ref{YL}, \ref{Y'}) satisfy those equations. The density has
the form
\begin{equation}
\sigma(x)=a_1(x)-\frac{1}{\lambda^2}a_2(x)
+\frac{\kappa}{\lambda^2} b(x).
\end{equation}
Plugging this form in the definition of $\kappa$, we obtain
\begin{equation}
\kappa =
\frac{2\int_{x_2}^{b}h(x)[a_1(x)-\frac{1}{\lambda^2}a_2(x)]dx}
{1-\frac{2}{\lambda^2}\int_{x_2}^{b}h(x)b(x)dx}
\end{equation}
which, expanded in powers of $\lambda$, is given by
\begin{equation}
\kappa = 2(g-h_1)+\lambda^2 \mu_1+O(\lambda^4).
\end{equation}
where
\begin{equation}\label{mu}
\mu_1\equiv \frac{1}{\beta}[-\alpha_1+2(g-h_1)]
\end{equation}
and
\begin{equation}
\begin{array}{ll}
\alpha_1\equiv &\displaystyle{ \frac{2}{\gamma_{12}}\int_{x_2}^{b}dx
\left(\frac{V'(x)}{f'(x)}\right)' h(x)}\\
\beta\equiv &
\displaystyle{\frac{2}{\gamma_{12}}\int_{x_2}^{b}dx
\left(\frac{h'(x)}{f'(x)}\right)' h(x)}.
\end{array}
\end{equation}
Using the expansion for $\kappa$, given by Eq.~(48), we obtain in
the limit $\lambda\rightarrow 0$ from Eq.~(42)
\begin{equation}\label{Y}
Y(x)=V(x)+\mu_1 h(x).
\end{equation}
The free energy can then be written as
\begin{equation}\label{Fsp}
F(x_1,x_2)=H_1+\int_{x_2}^{b}dx \sigma
(x)\left[V(x)-f(x)\frac{Y'(x)}{f'(x)}\right].
\end{equation}
There are two additional constraints that were not included in the
variational scheme and will be enforced directly,
\begin{equation}
\sigma(x_2)\ge 0; \;\;\; x_2 > x_1.
\end{equation}

The conductance distribution now becomes
\begin{equation}\label{pg}
P(g)=\int_{0}^b dx_1 \int_{x_1}^{b}dx_2 e^{-F(x_1,x_2;g)}
\delta(Y'(x_2)),
\end{equation}
with equations (\ref{V}), (\ref{f}), (\ref{sigma}) and (\ref{Y})
defining $V(x)$, $f(x)$, $\sigma(x)$, and $Y(x)$, respectively.

\subsection{The free energy}

Let us write $F(x_1,x_2)=H_1(x_1,x_2)+F_2(x_2)$ and define
\begin{equation}
W(x)=Y'(x)/f'(x).
\end{equation}
Then $\sigma(x)=W'(x)/\gamma_{12}$ and
\begin{equation}
F_2=\frac{1}{\gamma_{12}}\left[\int_{x_2}^{b}dx W'(x)V(x)-
\int_{x_2}^{b}dx W'(x)f(x)W(x)\right].
\end{equation}
On the other hand, defining
\begin{equation}
\Phi(x)=V(x)-f(x)W(x)
\end{equation}
and using partial integration, we get
\begin{equation}
F_2=-\frac{1}{\gamma_{12}}\int_{x_2}^{b}dx W(x)
\left[V'(x)-f'(x)W(x)- f(x)W'(x)\right],
\end{equation}
where we have neglected an irrelevant term
$\Phi(b)W(b)/\gamma_{12}$ independent of $x_2$ and we have used
$W(x_2)=0$. Using $f'(x)W(x)=Y'(x)$ and $V'(x)-Y'(x)=-\mu_1h'(x)$,
we rewrite the above as
\begin{equation}\label{F21}
F_2=\frac{\mu_1}{\gamma_{12}}\int_{x_2}^{b}dx h'(x)W(x)+
\frac{1}{\gamma_{12}}\int_{x_2}^{b}dx f(x)W'(x)W(x).
\end{equation}
The two alternate expressions for $F_2$ can now be combined to
obtain
\begin{equation}
\begin{array}{rl}
\int_{x_2}^{b}dx f(x)W'(x)W(x)=& \displaystyle{\frac{1}{2}\int_{x_2}^{b}dx
W'(x)V(x)}\\
-&\displaystyle{ \frac{\mu_1}{2}\int_{x_2}^{b}dx h'(x)W(x)+ C,}
\end{array}
\end{equation}
where $C$ is a constant. Plugging this back to Eq.~(\ref{F21}), we
obtain
\begin{equation}
F_2=\frac{1}{2\gamma_{12}}\int_{x_2}^{b}dx W'(x)V(x)+
\frac{\mu_1}{2\gamma_{12}}\int_{x_2}^{b}dx h'(x)W(x).
\end{equation}
We can again use partial integration to rewrite the first term as
an integral over $V'(x)W(x)$, using again the fact that
$W(x_2)=0$. Then the two terms can be combined to obtain
\begin{equation}\label{F2}
F_2=-\frac{1}{2\gamma_{12}}\int_{x_2}^{b}\frac{dx}{f'(x)}\left[
V'^2(x)-\mu^2_1 h'^2(x)\right].
\end{equation}
We define
\begin{equation}
\begin{array}{ll}
&\displaystyle{J_1=\int_{x_2}^b dx\frac{V'^2(x)}{f'(x)}; \;\;\; J_2=\int_{x_2}^b
dx\frac{V'(x)h'(x)}{f'(x)}};\\
&\displaystyle{J_3=\int_{x_2}^b dx
\frac{h'^2(x)}{f'(x)}}.
\end{array}
\end{equation}
Then
\begin{equation}
F_2=-\frac{1}{2\gamma_{12}}[J_1-\mu^2_1J_3]; \;\;\;
\mu_1=-\frac{V'(x_2)}{h'(x_2)}
\end{equation}

\subsection{The constraints}

We already have one constraint $Y'(x_2)=0$. We also demand
$\sigma(x_2)\ge 0$ which requires
\begin{equation}
Y''(x_2)\ge 0.
\end{equation}
This defines $x_{2min}$. Also, from Eq.~(\ref{mu}),
\begin{equation}
g-1/\cosh^2 x_1=\frac{1}{2}(\alpha_1+\mu_1\beta)\equiv g_0,
\end{equation}
or
\begin{equation}
x_1=\cosh^{-1}\frac{1}{\sqrt{g-g_0}}.
\end{equation}
On the other hand defining $W_1(x)=V'(x)/f'(x)$ and
$W_2(x)=h'(x)/f'(x)$, it is easy to see that
\begin{equation}
g_0=\int_{x_2}^b dx h(x)\sigma(x).
\end{equation}
Using partial integration and the fact that $Y'(x_2)=0$, we can
also rewrite
\begin{equation}
g_0=-\frac{1}{\gamma_{12}}[J_2+\mu_1J_3].
\end{equation}

\subsection{Validity of the approximations}

We started with the assumption that while $\langle x_1\rangle\gg
1$ in the insulating regime, the difference $\langle
x_2-x_1\rangle\ll \langle x_1\rangle$. It is therefore important
to estimate the difference from the above results. We will use
saddle points of the free energy
\begin{equation}
\begin{array}{ll}
F(x_1,x_2)\approx &V(x_1)-\ln (x^2_2-x^2_1)+ F_2(x_2);\\
~&V(x_1)= \Gamma x^2_1-x_1-\ln x_1
\end{array}
\end{equation}
where $F_2$ is given in Eq.~(\ref{F2}). Let us define $s= x_2-x_1$
and $\zeta=x_1-1/2\Gamma$. We assume $s \ll x_1$ and $\zeta \ll
x_1$. The saddle point solutions for $s$ and $\zeta$ are obtained
from $\partial F/\partial \zeta=0$ and $\partial F/\partial s=0$.
Using chain rule to write the partial derivatives in terms of
$\partial F/\partial x_1$ and $\partial F/\partial x_2$ we obtain
\begin{equation}\label{ds}
V'(x_1)-\frac{2}{x_2+x_1}+F'_2(x_2)=0; \;\;\;
-\frac{1}{s}-\frac{1}{x_2+x_1}+F'_2(x_2)=0
\end{equation}
where prime denotes derivatives with respect to the arguments.
Combining the two gives
\begin{equation}
-\zeta s\approx \frac{1}{2\Gamma}, \;\;\; \Gamma \ll 1.
\end{equation}
From the definition of $F_2$ we have
\begin{equation}\label{F2'}
F'_2(x_2)=-\frac{1}{2\gamma_{12}}\left[ \frac{\partial
J_1}{\partial x_2} -\mu^2_1 \frac{\partial J_3}{\partial
x_2}\right]+\frac{\mu_1}{\gamma_{12}} \frac{\partial
\mu_1}{\partial x_2}J_3.
\end{equation}
The integrals $J_1$ and $J_3$ depend on $x_2$ only via the lower
limits of the integrals. Derivatives w.r.t. $x_2$ are simply the
negatives of the integrands evaluated at the lower limit. Using
definition of $\mu_1$ this gives the first term in Eq.~(\ref{F2'})
equal to zero, leaving $F'_2(x_2)=\frac{\mu_1}{\gamma_{12}}
\frac{\partial \mu_1}{\partial x_2}J_3.$ Using $h(x)\approx
4e^{-2x}$ and $1/f'(x)\approx x/k$ for $x/k \ll 1$, we get
\begin{equation}
J_3 \approx \frac{1}{k}\left[16 x_2e^{-4x_2}+4 e^{-4x_2}\right].
\end{equation}
Then
\begin{equation}
F'_2(x_2)\approx
\Gamma^2x^3_2-\Gamma_3x^2_2+\frac{1}{4}(1+\Gamma_2)x_2
+\frac{9}{16}(1+\Gamma_1),
\end{equation}
where $\Gamma_3=\Gamma(1-3\Gamma/4)$, $\Gamma_2=-6\Gamma
+\Gamma^2/2)$ and $\Gamma_1=-\Gamma/3$. We neglect $1/(x_2+x_1)$
compared to the other terms in $F'(s)=0$ (Eq.~(\ref{ds})) and
expand $F'_2(x_2)$ in Taylor series around $x_2=1/2\Gamma$. The
dominant term is $\frac{1}{2}(\zeta+s)^2\Gamma$ where we have used
$F'''_2(x_2=1/2\Gamma)=\Gamma$. Using $\zeta=-1/2\Gamma s$, we
finally obtain
\begin{equation}
\frac{\partial F}{\partial s}\approx \frac{1}{8\Gamma
s^2}-\frac{1}{2}+\frac{1}{2}\Gamma s^2 =0.
\end{equation}
Therefore, the saddle point solutions are given by
\begin{equation}\label{s}
s=-\zeta=\frac{1}{\sqrt{2\Gamma}}.
\end{equation}
Since $\langle x_1\rangle \sim 1/2\Gamma$, we confirm our
expectation that $\langle x_2-x_1\rangle \ll \langle x_1\rangle$.
The results are also consistent with our assumption that both $s$
and $\zeta$ are much smaller than $x_1$, so the free energy
calculations remain valid.

However, numerically we find that while $\langle x_1\rangle \sim
1/2\Gamma$, $s\sim 1$ independent of disorder. Thus our result
$s=\frac{1}{\sqrt{2\Gamma}} \ll \langle x_1\rangle $ is only
qualitatively correct. The fact that actual $s$ is much smaller
than what we find is related to the inaccuracy in our evaluation
of the density. Indeed, we can obtain the density directly from
Eq.~(\ref{sigma}). In the limit $x \gg 1$, $\gamma_{12} \ll 1$ we
find
\begin{equation}
\begin{array}{ll}
\sigma (x)&\approx
\frac{1}{\gamma_{12}}\Big[2\Gamma-\frac{k(2\Gamma k+1)}{(x+k)^2}\\
~& +
2V'(x_2)\left(\frac{x}{x+k}-\frac{k}{2(x+k)^2}\right)e^{-2(x-x_2)}\Big]
\end{array}
\end{equation}
where we have used $k=2/\gamma_{12}$ (Eq~(\ref{f})).  In the limit
$x_2\ll k$ but $x \gtrsim x_2$, the density simplifies to
\begin{equation}\label{sigapp}
\sigma (x)\approx 2\Gamma x.
\end{equation}
The linear $x$-dependence as well as the $\Gamma$ dependence
agrees with numerical results. However, the slope turns out to be
too large. This is possibly the consequence of our simplification
of the Hamiltonian Eq.~(\ref{ham}) to Eq.~(\ref{happ}), where all
the interaction terms were neglected except for the one between
the first and the second levels. As shown in
[\onlinecite{mu-wo-99}], it should be possible to obtain an
integral equation for the saddle point density which can then be
solved at least approximately.

As we will show, the actual density of the levels play a minor
role in the distribution $P(g)$, which is dominated by the first
few levels. Therefore our results will be qualitatively correct,
although there would be quantitative discrepancies due to the
difference in the density.

Note that in the opposite limit $x_2\gg k$, $x \gg x_2$, the
density becomes
\begin{equation}
\sigma (x)\approx 2\Gamma /\gamma_{12}.
\end{equation}
This corresponds to a uniform average spacing $s=\langle x_{a+1} -
x_a \rangle$ of eigenvalues of order unity ($L=L_z$), compared to
the uniform spacing $s\approx L_z/\xi$ in Q1D. In contrast, 3D
metals are similar to Q1D metals having uniform $\sigma(x)$
extending down to $x=0$ and $s\sim L_z/L^2$. The opening of a gap
in the spectrum of Lyapunov exponents $\nu_n \equiv \langle x_n
\rangle/L_z\sim 1/\xi$ may be considered as the signature of the
Anderson transition.

\section{Results and discussions}

With the above caveat in mind, the saddle point free energy
$F_{sp}(x_1,x_2;g)$ has the form (Eq.~(\ref{Fsp}))
\begin{equation}\label{res-1}
F_{sp}(x_1,x_2) =  H_1 - \frac{1}{2\gamma_{12}}\int_{x_2}^b
\frac{dx}{f'(x)} [V'^2(x)-\mu^2_1 h'^2(x)],
\end{equation}
where primes denote $x$-derivatives. Eq~(\ref{pg}) can then be
rewritten as
\begin{equation}\label{model}
P(\ln g)\propto g \int_{x_{2min}}^{b}dx_2
e^{-F_{sp}(x_1,x_2;g)}e^{-2(x_2-x_1)},
\end{equation}
where the integration over $x_1$ is eliminated by a constraint
arising from the minimization of the free energy:
\begin{equation}\label{res-2}
\begin{array}{ll}
x_1 = &\cosh^{-1}[1/\sqrt{g - g_0}]; \\
~&~\\
~&g_0  =
-\frac{1}{\gamma_{12}}\int_{x_2}^b
\frac{dx}{f'(x)}h'(x)[V'(x)+\mu_1h'(x)].
\end{array}
\end{equation}
The lower limit $x_{2min}$ is the larger of the additional
constraints imposed by the conditions $\sigma(x_2)\ge 0$ and $x_2
> x_1 \ge 0$, $x_1 $ real.

\subsection{Analytical model}

It is instructive to consider first a simple approximate solution
of Eq.~(\ref{model}), which is dominated by the lower limit of the
integral.  To a good approximation, $g_0$ is negligible compared
to $g$ in the insulating limit, and $x_1\approx \frac{1}{2}\ln
(4/g)$. The condition $\sigma(x_2) \ge 0$ or equivalently $Y''(x)
\ge 0$ gives $x_{2min}$ from the condition  $Y''(x_{2min})= 0$.
This gives
\begin{equation}
x_{2min}\approx (1+\Gamma+\gamma_{12})/2\Gamma
\end{equation}
and hence $F_{sp}\approx H_1$. This immediately leads to
\begin{equation}\label{pa}
P(\ln g)\propto (4 x^2_{2min}-u^2)
e^{-\frac{\Gamma}{4}(\frac{1}{\Gamma}+u)^2}, \;\;\; u\equiv \ln
(g/4).
\end{equation}

\begin{figure}
\includegraphics[clip,width=0.35\textwidth]{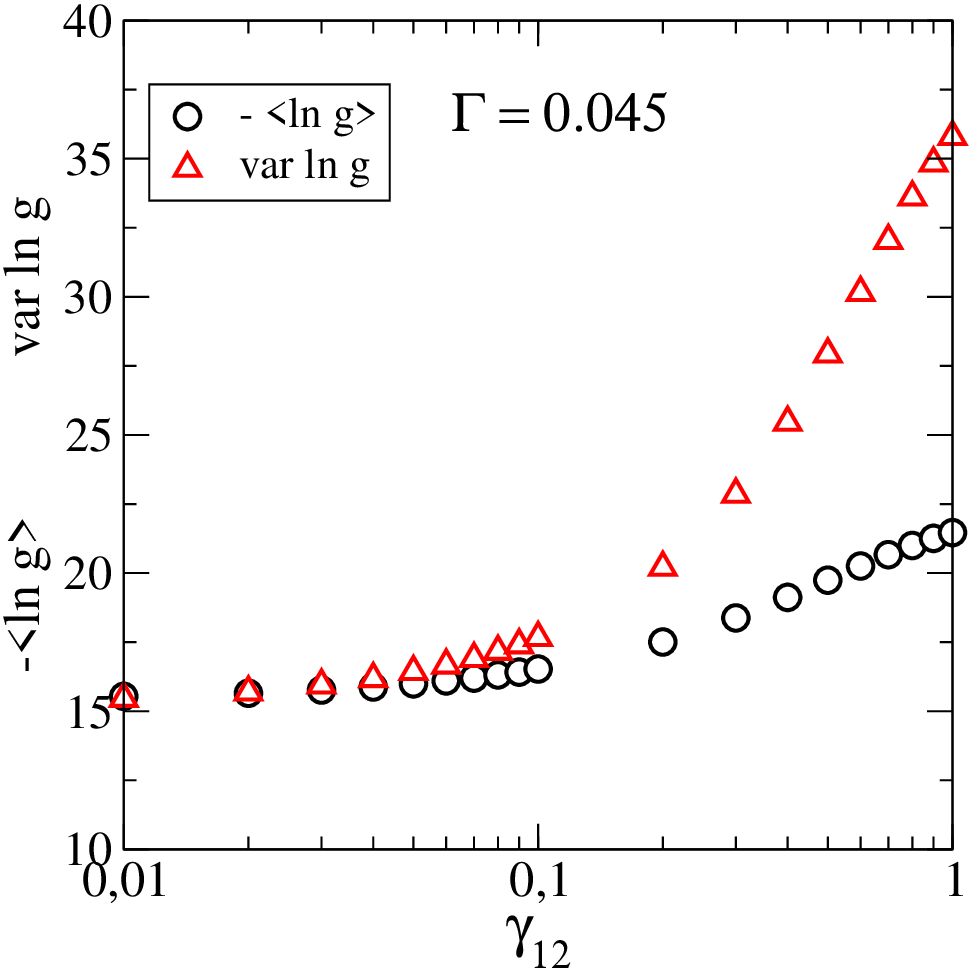}
\caption{Mean value $\langle\ln g\rangle$ and variance var $\ln g$
as a function of $\gamma_{12}$ in the strongly insulating regime,
calculated from analytical distribution (\ref{model}).
As expected, var $\ln g\approx -\langle\ln g\rangle$ for
$\gamma_{12}\ll 1$ but var $\ln g\approx -2\langle\ln g\rangle$
for $\gamma_{12}=1$. Note that both mean and variance depends only weakly
on $\gamma_{12}$ when $\gamma_{12}$ is small (which is always the case when
the system size $L$ is large).
} \label{var_gamma}
\end{figure}

We do a saddle point analysis of Eq.~(\ref{pa}) to obtain $\langle
\ln g\rangle$ and var($\ln g$)  as a function of $\Gamma$. In
order to illustrate the difference between Q1D and 3D insulators,
we will keep the general expressions without using the condition
$\gamma_{12} \ll 1$. The free energy can be written as
\begin{equation}
F_{approx}= \frac{\Gamma}{4}(x-\frac{1}{\Gamma})^2-\ln (2
x_{2min}-x); \;\;\; x=\ln(1/g).
\end{equation}
The saddle point solution of the mean $x_m\equiv \langle \ln g
\rangle$ is obtained from $F'(x_m)=0$ where the prime denotes
derivative with respect to $x$.  Denoting
\begin{equation}\label{spmean}
x_m=1/\Gamma -y
\end{equation}
this gives $y(y-\gamma_{12}/\Gamma)=2/\Gamma$, leading to
\begin{equation}\label{spy}
y=\frac{1}{2}\left[\sqrt{(\frac{\gamma_{12}}{\Gamma})^2+\frac{8}{\Gamma}}
-\frac{\gamma_{12}}{\Gamma}\right].
\end{equation}
The variance $\delta x$ can be estimated from $1/F''(x_m)$, giving
\begin{equation}\label{spvar}
\delta x \approx \frac{1}{\frac{\Gamma}{2}+\frac{\Gamma^2y^2}{4}}.
\end{equation}
In the limit $\gamma_{12}\rightarrow 0$ appropriate for our 3D
insulators, $y \rightarrow \sqrt{2/\Gamma}$. On the other hand in
the Q1D limit $\gamma_{12}^{Q1D}=1$, $y\rightarrow 2$. Thus
compared to the Q1D result $\langle \ln g \rangle^{Q1D}\approx
1/\Gamma$, the 3D result is shifted by $\sqrt{2/\Gamma}$.
Similarly, compared to the Q1D result var($\ln g) = 2/\Gamma$, the
3D insulators have a much sharper distribution, with half the
variance $1/\Gamma$. Both results agree with numerical data.
Although our model is not in general valid for
$\gamma_{12}\rightarrow 1$ because of our neglect of the
interaction terms, in the insulating limit the interaction terms
are negligible and it is useful to see how the mean and the
variance changes with
$\gamma_{12}$ as it is changed from the 3D
limiting value of zero to the Q1D limiting value of unity. Since
$\gamma_{12} \sim \xi/2L$, for a given disorder this will
correspond to starting from a cubic sample of width $L=L_z \gg
\xi$ (where $L_z$ is the length) and decreasing the width to $L <
\xi$ to reach the Q1D limit.

\begin{figure}
\includegraphics[clip,width=0.35\textwidth]{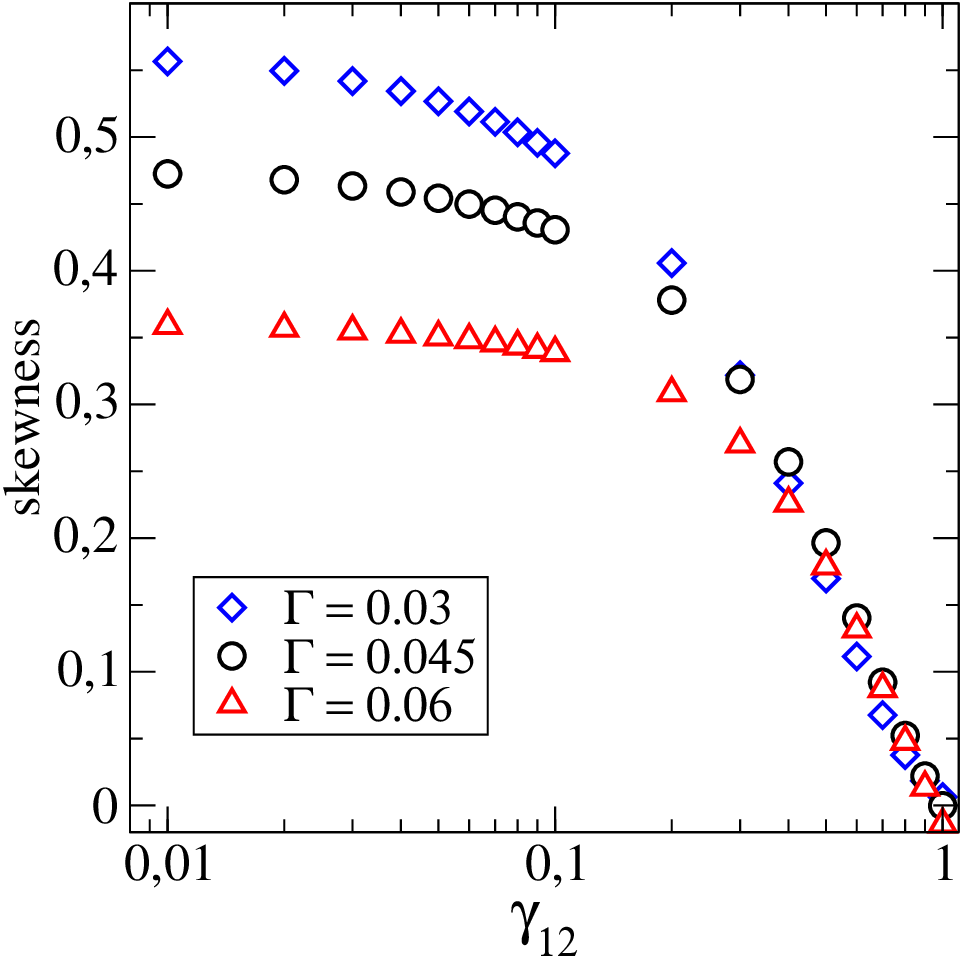}
\caption{Skewness as a function of $\gamma_{12}$ in the strongly
insulating regime. As expected, skewness is zero for
$\gamma_{12}=1$ (Q1D limit). } \label{skew_gamma}
\end{figure}

\begin{figure}[t]
\includegraphics[clip,width=0.35\textwidth]{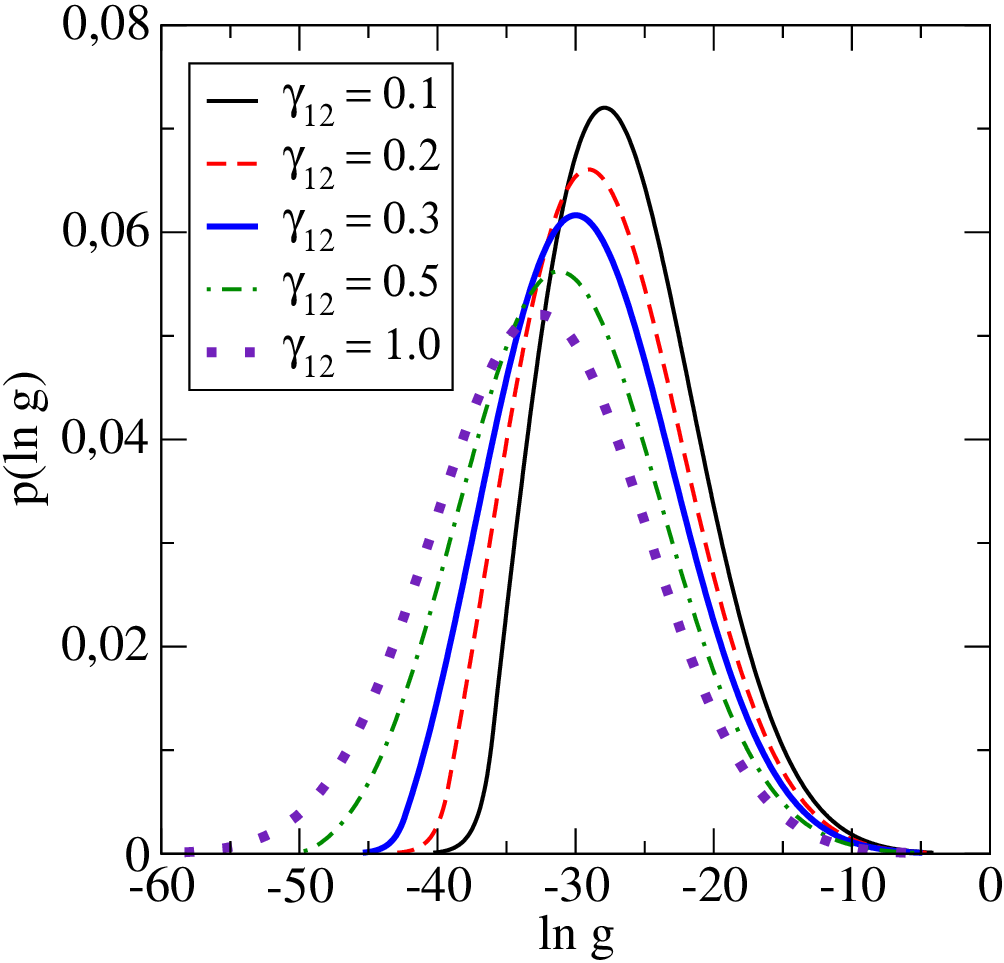}
\caption{$P(\ln g)$ obtained from analytical formulas
(Eqs.~\ref{res-1}, \ref{model} and \ref{res-2}) for $\Gamma=0.03$
 and various values of $\gamma_{12}$.}
\label{ann_plng}
\end{figure}

Figure \ref{var_gamma} shows how
the mean and the variance changes with $\gamma_{12}$ according to
Eqs.~(\ref{spmean}) and (\ref{spvar}).

\begin{figure}
\includegraphics[clip,width=0.4\textwidth]{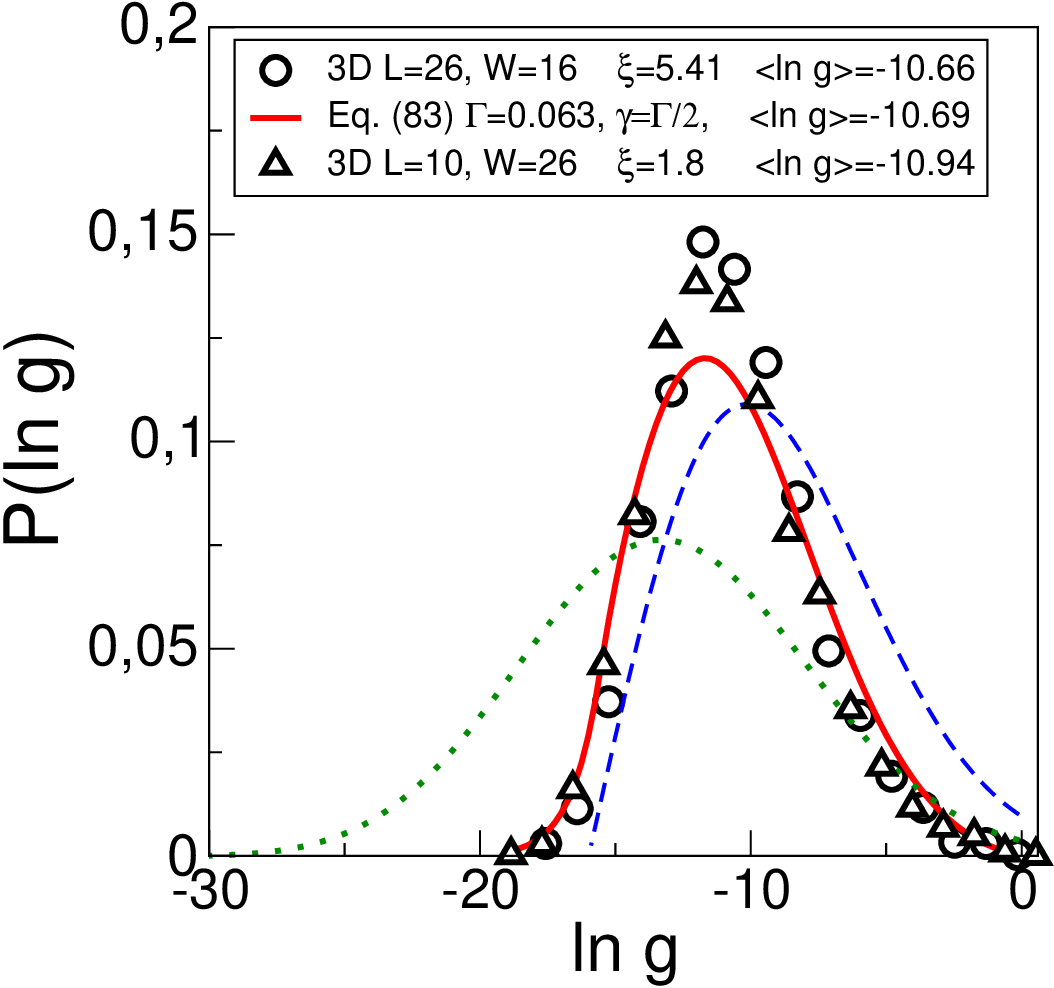}
\caption{Conductance distribution for 3D insulators obtained from
direct numerical simulation  for $W=16$, $L=26$ (circles), and
from Eq.~(\ref{model}) for $\Gamma=0.063$  and
$\gamma_{12}=\Gamma/2$ (solid line). Both have the same mean value
$\langle\ln g\rangle\approx-10.6$. Dashed and dotted lines show
Eq.~(\ref{pa}) with $\xxm=1/2\Gamma+3/4$ and $\xxm=1/\Gamma$,
respectively, with $\Gamma=0.063$. Shown are also numerical data
for $W=26$ and $L=10$ (triangles). Similar agreement is obtained
for other values of $\langle\ln g\rangle$ if $\Gamma$ is used as a
free parameter; see Ref. [\onlinecite{mmwk}] for the case
$\langle\ln g\rangle\approx-12.6$ fitted with $\Gamma=0.054$
\cite{noteG/g}.  } \label{mmwk}
\end{figure}

It is not possible to obtain a simple formula for the skewness
$\langle (\ln g- \langle \ln g\rangle)^3\rangle/[\langle (\ln g-
\langle \ln g\rangle)^2]^{3/2}$ except that in the limit
$\Gamma\rightarrow 0$ it approaches a number of order unity.
Direct evaluation of the quantity as a function of $\gamma_{12}$
is shown in Fig.~\ref{skew_gamma}, which shows that for a given
disorder, the skewness starts from zero in the Q1D limit, as is
well known, but saturates to a finite value (depending on
disorder) in the 3D limit. It shows that the distribution is never
log-normal for 3D insulators. It also shows that the distribution
$P(\ln g)$ is almost independent of $\gamma_{12}$ provided that
both $\gamma_{12}$ and $\Gamma$ are small. This explains why the
distribution shown in Fig.~\ref{mmwk} does not depend on the ratio
$\Gamma/\gamma$.

Finally, Fig.~\ref{ann_plng} shows how the entire distribution
$P(\ln g)$ changes as a function of $\gamma_{12}$, from a sharp,
skewed form for $\gamma_{12} \ll 1$ to a broad Gaussian form for
$\gamma_{12}= 1$.

\begin{figure}
\includegraphics[clip,width=0.4\textwidth]{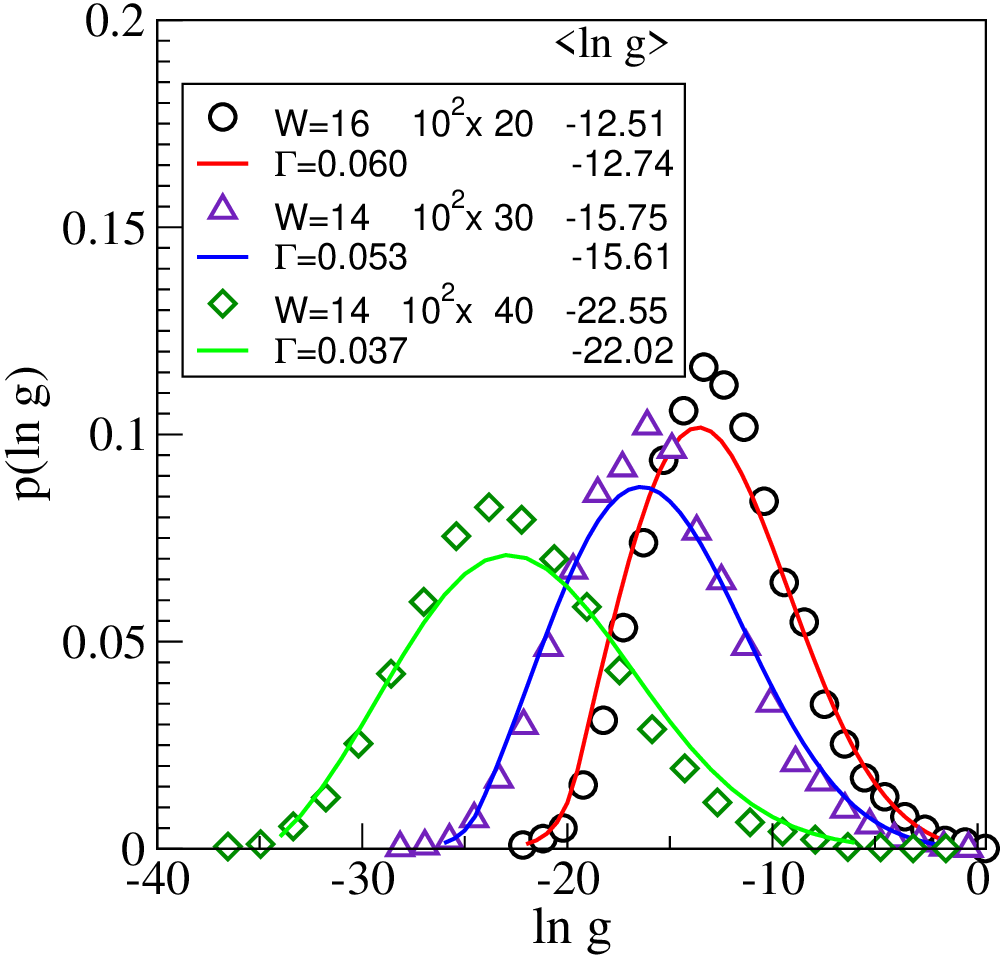}
\caption{Conductance distribution for insulating samples of the
size $L^2\times L_z$  ($L=10$)
In contrast to 3D system (fig.
\ref{mmwk}), parameter $\Gamma/\gamma_{12}< 1$. From numerical
data, we have $\Gamma=0.3$, $\gamma_{12}=0.22$ for $W=16$, and
$\Gamma=0.133~(0.102)$, $\gamma_{12}=0.32~(0.36)$ for systems with
$W=14$ and $L_z=30$ (40), respectively.
Solid lines show analytical result, Eq. (\ref{model})
with $\gamma_{12}$ and $\langle\ln g\rangle$ as given form numerical data.
} \label{mmwk-b}
\end{figure}

\subsection{Comparison with numerical data}

When comparing our theoretical prediction, Eq. (\ref{model}), with
numerical data, we distinguish two cases: (1)  for 3D systems,
both $\Gamma$ and $\gamma_{12}$ decreases $\sim 1/L$, while the
ratio $\Gamma/\gamma_{12}$ does not depend on $L$. (2) for systems
$L^2\times L_z$, $\Gamma\sim 1/L_z$ while $\gamma_{12}\sim 1/L$
does not depend on $L_z$ (see fig. \ref{3D_t_cp}). Therefore,
contrary to 3D geometry, $\Gamma/\gamma_{12}\ll 1$ for $L_z\gg L$.
This  different behavior of parameters $\Gamma$ and $\gamma_{12}$
explains difference between  the shape of $p(\ln g)$  in 3D and
Q1D strongly disordered systems, shown in figures \ref{mmwk} and
\ref{mmwk-b}.

Figure \ref{mmwk} shows Eq.~(\ref{pa}) compared with the results
from direct integration of Eq.~(\ref{model}), both compared with
numerical results based on Eq.~(\ref{AndHam}). For the analytic
curves, we chose $\Gamma=0.063$ and $\gamma_{12}=\Gamma/2$ to have
the same $\langle\ln g\rangle$ as in the numerical case
\cite{noteG/g}. Note that using the Q1D result
$\gamma_{12}^{Q1D}=1$ gives $x_{2min}\approx 1/\Gamma$, leading to
a log-normal distribution (see dotted line in Fig.~(\ref{mmwk})).
As shown in [\onlinecite{mmwk}], variance and skewness calculated
from direct integration of Eq.~(\ref{model}) compares well with
numerical results, consistent with saddle point results from
Eq.~(\ref{pa}).

Figure \ref{mmwk-b} compares theoretical formula, Eq.
(\ref{model}) with numerical data for  $p(\ln g)$  for insulating
samples $L^2\times L_z$. While $\Gamma$ is small, decreasing as
$\sim 1/L_z$, $\gamma_{12}$ does not depend on $L_z$ and is
constant for fixed $L$. Consequently, ratio $\gamma_{12}/
\Gamma\sim L_z/L$ increases with increasing $L_z$.

\smallskip

Both figures \ref{mmwk} and \ref{mmwk-b} show qualitative agreement with
numerical data and theoretical model. Quantitative differences
between Eq.~(\ref{model}) and numerical results  have origin in
our simplified model
Eq.~(\ref{simple}),  which still overestimates the strength of the
interaction for higher channels.

\medskip

\begin{figure}[t]
\includegraphics[clip,width=0.35\textwidth]{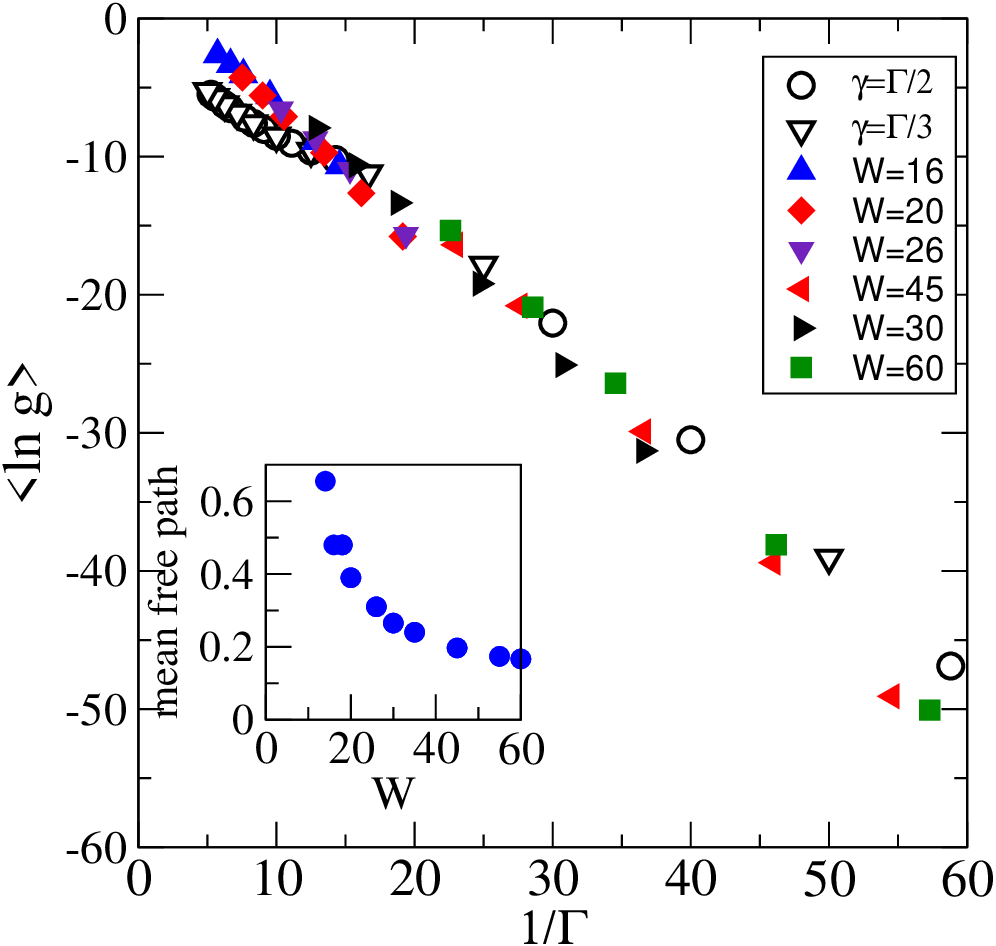}
\caption{Mean value $\langle\ln g\rangle$ as a function of
$1/\Gamma$. 
Open symbols: analytical result, Eq. (\ref{model}). Data confirm 
linear dependence 
$\langle\ln g \rangle\propto 1/\Gamma$ for small $\Gamma$ (strong disorder).
For smaller disorder (larger $\Gamma$)
analytical model  is less accurate.
Data confirm that $\langle \ln g\rangle$ does not depend on
$\gamma_{12}$ provided that both $\Gamma$ and $\gamma_{12}$ are small.
Full symbols: numerical data with $\Gamma^{-1}=LK_{11}/\ell$. 
We estimate $\Gamma$ for a given disorder using the mean free path $\ell$ 
obtained from data in fig.~19a ($\ell(W)$ is shown in inset) 
and the relation $K_{11}L=x_1\ell$.  
Numerical 
$\langle\ln g\rangle$ is an unambiguous function of $\Gamma$, as required in 
the analytical model. Small  deviations are due to finite size effects 
which affect actual values of parameters $x_1$ and $K_{11}$.
}
\label{G-lng}
\end{figure}

It is important to note that in both figures \ref{mmwk}  and \ref{mmwk-b} we compared
numerical data with theoretical model with the same mean
$\langle\ln g\rangle$. This is consistent with scaling theory of
localization since there is only one parameter - for instance
$\langle\ln g\rangle$ - which determines $p(\ln g)$ completely.
In order to make sure that the analytical model has  the same 
$\langle\ln g\rangle$ as the numerical data for a given disorder,
we used $\Gamma$ as a free fitting parameter. Of course we could use Eq.~(20)
to obtain $\Gamma$ independently for a given disorder. However, in order 
to do that  
we will need a good estimate of the mean free path $\ell$. This is difficult
in the strongly disordered regime because the mean free path defined as the 
decay length of the single particle Green's function is actually 
smaller than the lattice spacing in the 
strongly disordered regime \cite{economou}, and our numerical model does
not allow us to obtain such small lengths with good accuracy. 
While independent calculations of the mean free 
path are available for cubic systems below critical disorder \cite{economou}
(e.g. $\ell=0.234$ for $W=15$ in the isotropic case), there is no data available
in the insulating regime. We therefor use $\Gamma$ as a free parameter. 
Nevertheless, as a consistency check, we estimate $\ell$ from  
a plot of $K_{11}L$ vs $x_1$ (fig.~19a), where the slope 
should  give the mean free path (equivalently, we could  identify
$\Gamma^{-1}$ with numerical value of $x_1$).
The results are plotted as the 
inset in fig. \ref{G-lng} to show the mean free path as a function of disorder. 
Using this result, we can estimate the value of $\Gamma$ corresponding to 
the disorder $W=16$ used in fig.~25. We find that $\ml(W=16)\approx 0.48$ 
and $\Gamma=0.070$, which is 
close to the fitting value $0.063$. This shows that while we can not obtain
$\Gamma$ accurately enough in our present numerical scheme, the fitted
values are consistent with our crude estimates. 
As a further consistency check, we use the above 
estimate of the mean free path 
to plot in fig.~27 the $\Gamma$ dependence of $\langle\ln g\rangle$. It shows 
that $\langle\ln g\rangle$ is 
indeed an unambiguous function of $\Gamma$, as required by the theory.        

Finally, we have checked the effects of fluctuations of $k_{11}$ on $P(\ln g)$
by integrating the conductance distribution in fig.~25 over the distribution $P(k_{11})$
(fig. \ref{loc_a}). We find that the effects are negligible.

\section{Beyond the insulating limit }

\begin{figure}
\includegraphics[clip,width=0.35\textwidth]{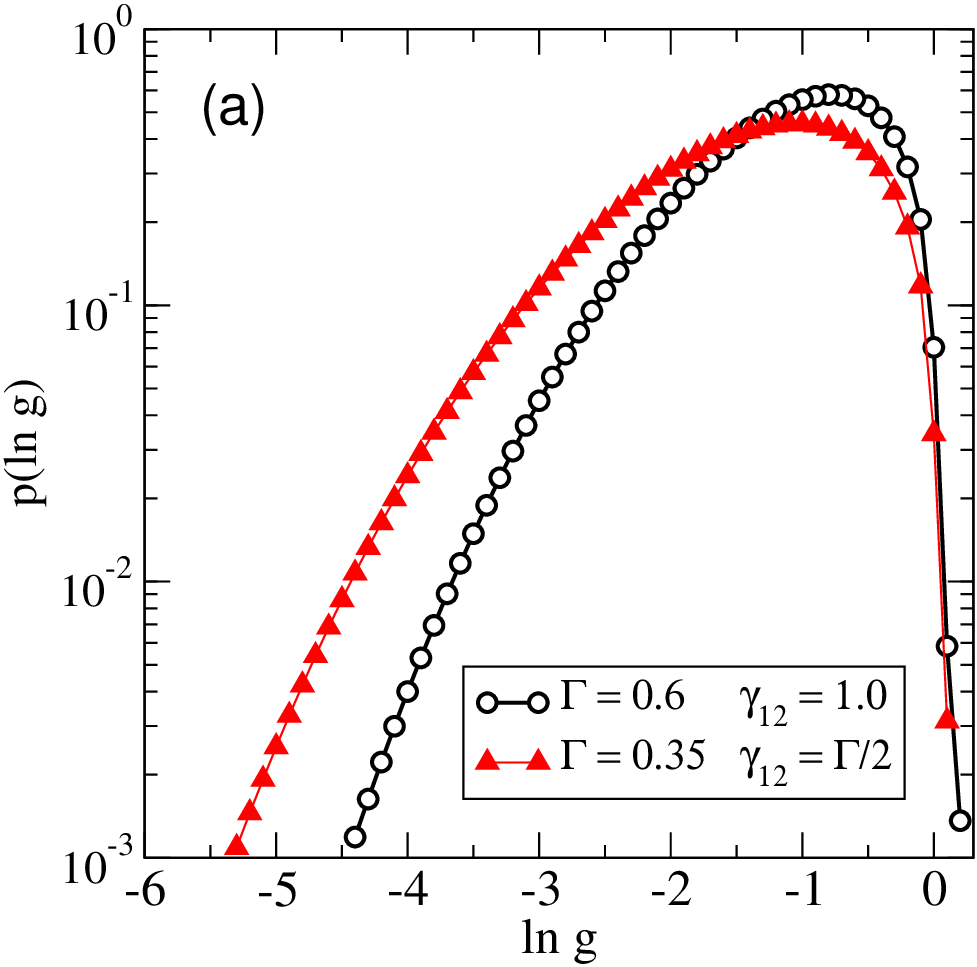}
\includegraphics[clip,width=0.35\textwidth]{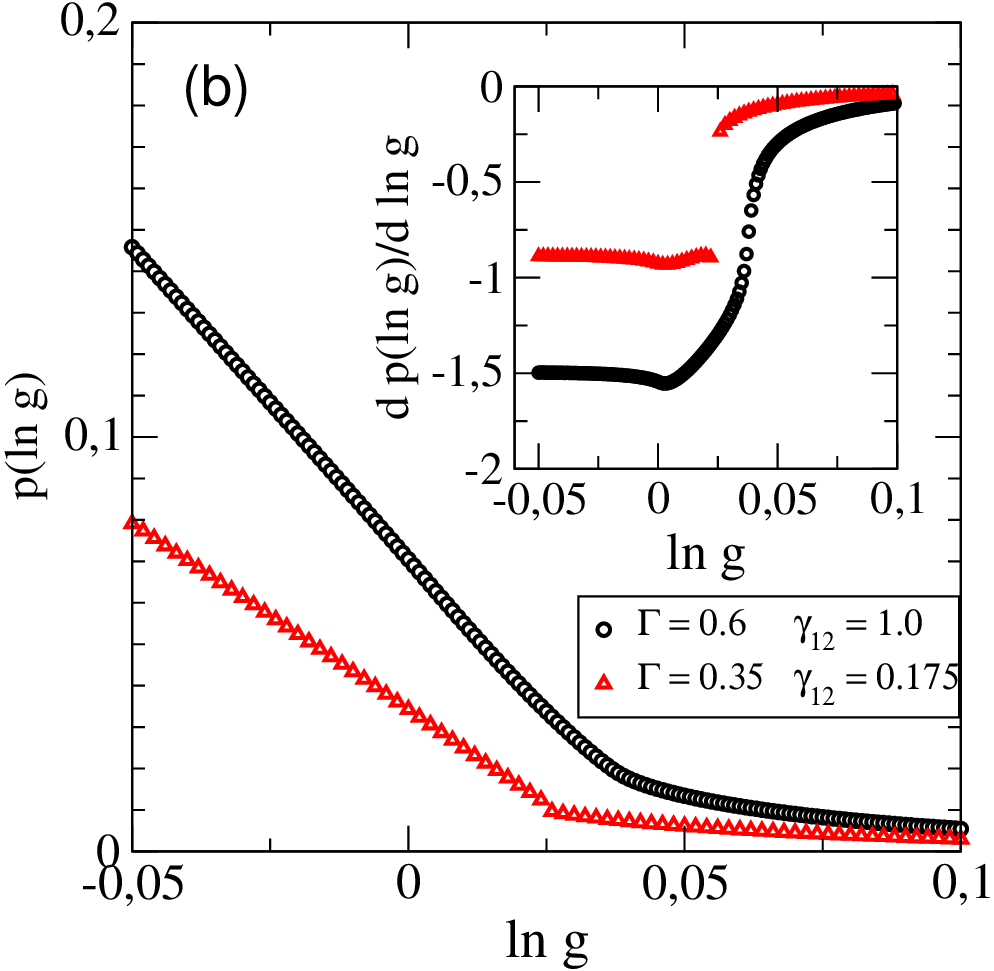}
\caption{(a) Probability distribution $P(\ln g)$ for $\Gamma\sim
1$ where critical regime is expected. The distribution agrees
qualitatively with numerical data for the critical regime. Figure
(b) shows detail of the distribution shown for $\Gamma=0.35$,
$\gamma_{12}=\Gamma/2$ and for $\Gamma=0.6$, $\gamma_{12}=1$. It
shows that there is indeed a non-analyticity in the distribution
close to $\ln g=0$ with position of the non-analyticity at $\ln g
>0$, in agreement with analytical results \cite{mu-wo-ga-go-03}. }
\label{plng_g1}
\end{figure}

\begin{figure}
\includegraphics[clip,width=0.35\textwidth]{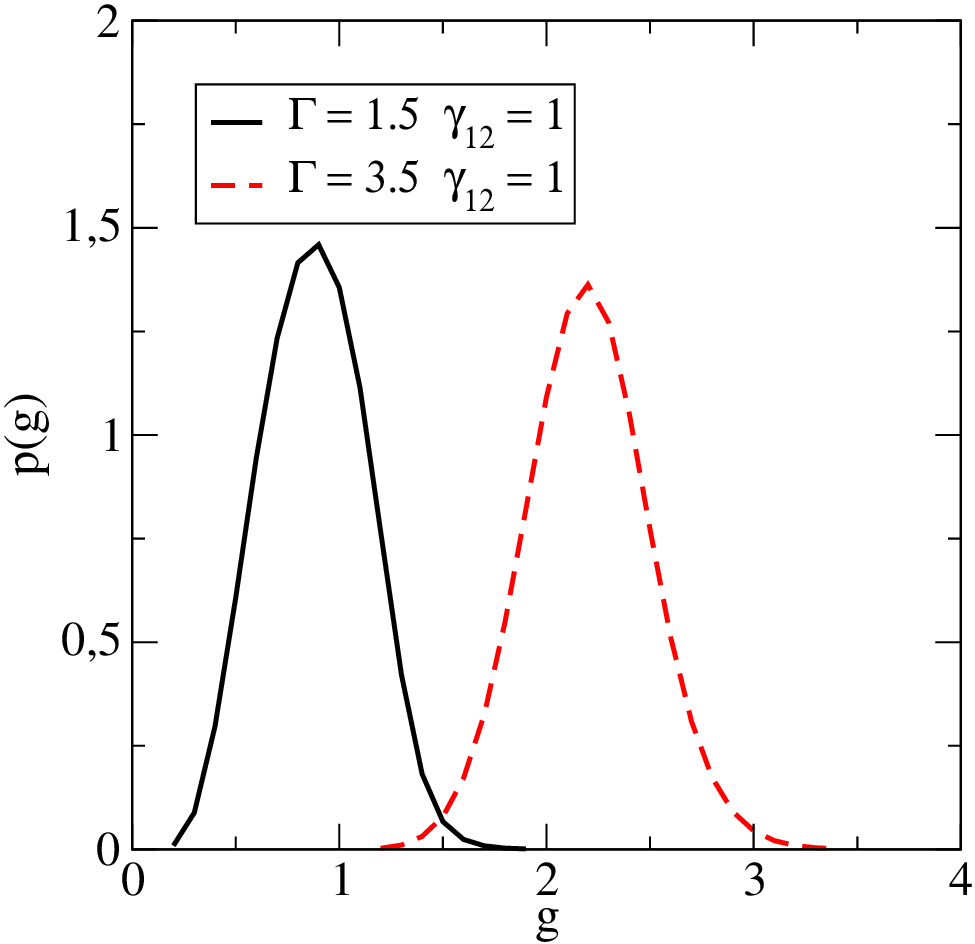}
\caption{Analytic result for the metallic regime ($\Gamma > 1$ and
$\gamma_{12}=1$). Data confirm that the distribution $P(g)$ is
Gaussian with variance var $g=0.064$ for $\Gamma=1.5$ and var
$g=0.09$ for $\Gamma=3.5$ which is  comparable  to the UCF value.}
\label{ann_metal}
\end{figure}

From numerical simulations \cite{MK-PM,markos1} we know that the
critical regime in 3D is also dominated by only a few eigenvalues
$x_i \gg 1$. However, since $\gamma_{12}$ is neither $0$ nor $2$,
it seems that we may not be able to use the free fermion
($\eta=0$) solution of Eq.~(\ref{ch}) to obtain the distribution
of the transmission levels. However, as shown in
[\onlinecite{caselle}], the solution is independent of the
strength of the interaction $\eta$ in the strong disorder regime
characterized by $x_i \gg 1$. This means that our solutions might
be used, albeit only qualitatively, even near the critical regime.
We show in Fig.~(\ref{plng_g1}) the distribution $P(\ln g)$ for
$\Gamma\sim 1$ which is expected to be near the critical regime.
It agrees qualitatively well with numerical results at the
critical point, including a discontinuity in the slope near $g=1$.
It is known from analysis of the Q1D systems \cite{mu-wo-ga-go-03}
that separating out an additional level helps to study the
non-analyticity near $g=1$. We therefore expect to obtain better
results near the critical regime by separating out an additional
level.

Finally, we show in Fig.~\ref{ann_metal} the distribution $P(g)$
for $\Gamma>1$ and for $\gamma_{12}=1$ which corresponds to the
metallic regime. Although we do not expect that our approximate
formula works quantitatively for the metallic regime,
Eq.~(\ref{model}) gives, for this choice of the parameters, a
Gaussian distribution of the conductance. Also the width  of the
distribution qualitatively agrees with the universal conductance
fluctuations \cite{ucf} in this regime. This shows that our simple
model already captures the essential qualitative features at all
strengths of disorder.

\section{Summary and conclusion}

We systematically analyzed the length and disorder dependence of
the matrix $K$ to  check if the generalized DMPK equation proposed
in [\onlinecite{mu-go-02}] is valid in three dimensional systems
at all strengths of disorder.
%Fluctuations of the elements at strong disorder may change the actual values of the parameters of the matrix $K$ but does not change their size or disorder dependence.

We studied the matrix $K$ in detail. The goal was to test the
assumptions on which the generalized DMPK equation was derived and
to construct a simple analytically tractable model for $K$ which
captures all the important qualitative features. In particular,
since Q1D systems have been studied in great detail, we looked for
any major qualitative differences in the structure of $K$ between
Q1D and 3D systems.

\begin{figure}
\includegraphics[clip,width=0.35\textwidth]{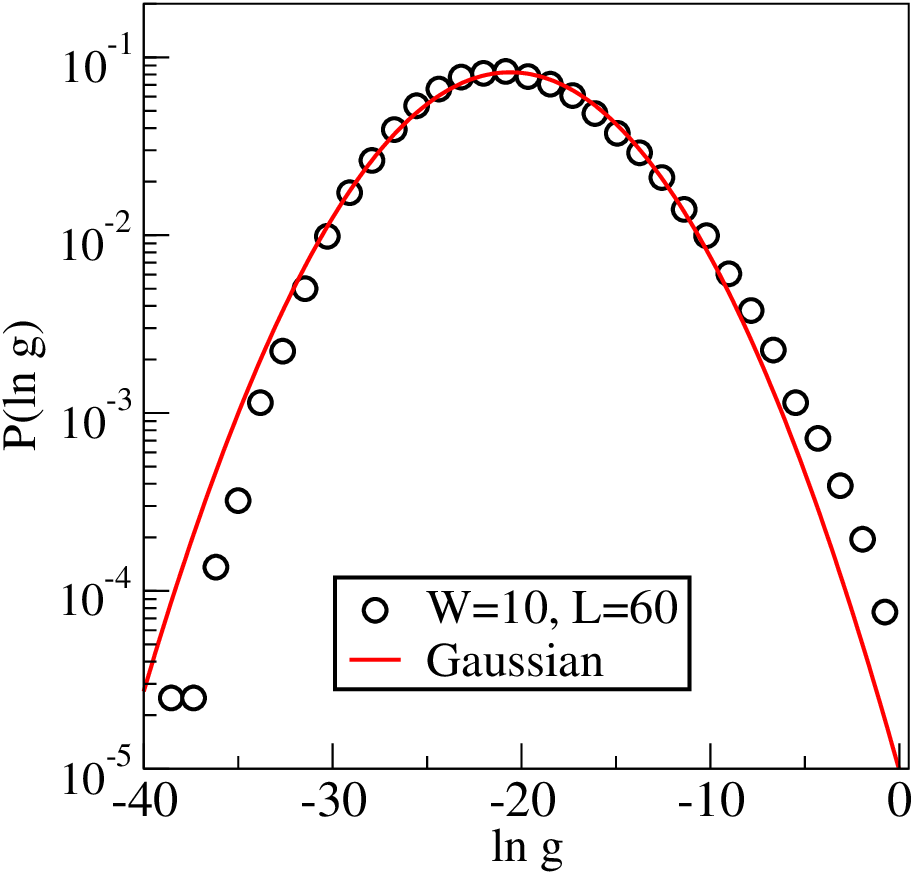}
\caption{Typical form of the distribution $p(\ln g)$ in 2D
disordered systems compared with Gaussian distribution with the
same mean value and variance. $\langle\ln g\rangle=-20.6$, var
$\ln g=23.5$ and skewness is 0.251. Figure indicates that the
deviation from the Gaussian distribution exists in 2D as well. }
\label{2D}
\end{figure}

We find that to a good approximation, the generalized DMPK
equation remains \textit{qualitatively} valid for any disorder. We
also conclude that to a good approximation, we can use only two
parameters, $K_{aa}\approx K_{11}$ and $\gamma_{ab}\approx
\gamma_{12}$, to characterize the qualitative changes in transport
at different strengths of disorder in different dimensions. 
We find that although fluctuations in $k_{11}$ at strong disorder are large (non
self-averaging), the effect of these fluctuations on $P(\ln g)$ is negligible.
More importantly, we do not have an independent way to estimate  
the mean free path to obtain 
$\Gamma=\ell/(L_zK_{11})$, but all qualitative features of the entire
distribution $P(\ln g)$ is obtained correctly once an effective
$\Gamma$ is used as a free parameter. We also find how these
parameters depend on disorder and show their unambiguous
dependence on the localization length. This is important since it
indicates that the introduction of new parameters does not
necessarily invalidate the single parameter scaling theory of
localization.

We have also shown that the matrix $K$ contains information about
the Anderson transition. The scaling of the parameters $LK_{11}$
or $\gamma_{12}$ clearly identifies the critical point which
agrees with numerical results.

We then concentrate on the strong disorder limit where our
numerical results allowed us to construct a simple
\textsl{one-parameter model} of the matrix $K$, containing
$\Gamma=\ell/K_{11}L_z$ and $\gamma_{12}=2K_{12}/K_{11}$, with
$\Gamma/\gamma_{12}=2$. By varying $\gamma_{12}$, we show how one
can go from a Q1D ($\xi \gg L$) to a truly 3D ($\xi \ll L$) system
in the insulating regime, which clearly shows the difference
between a Q1D and a 3D insulator. We then use the model to obtain
the full distribution $P(g)$ which agrees qualitatively with
numerical results.

It is indeed remarkable that even though the generalized DMPK
equation (\ref{genDMPK}) neglects fluctuations in $k_{ab}$ and the
model Eq.~(19) neglects the index dependence of $K_{ab}$, the
theory still captures all the essential features of length,
disorder as well as dimensionality dependence of the entire
conductance distribution and provides in particular a simple
understanding of the 3D distribution at strong disorder, which is
qualitatively different from a log-normal distribution in Q1D. At
the same time, our numerical studies suggest that having an 
independent estimate of the mean free path
could provide a more quantitative description of the
conductance distribution in 3D at all disorder.

We emphasize that there are large differences between  Q1D and
higher dimensions. In Q1D defined in
[\onlinecite{mu-wo-99,go-mu-wo-02,mu-wo-ga-go-03,dmpk}], disorder
is always \textsl{weak} enough to assure that  the localization
length $\xi \gg L$ where $L$ is the transverse dimension. The Q1D
insulator corresponds to the \textsl{weakly disordered} systems of
length $L_z \gg \xi$. It is this length-induced insulating
behavior that is described by the DMPK equation. This is different
from localization in 3D which occurs at \textsl{strong} disorder,
where $\xi$ is much less than both $L_z$ and $L$. This difference
is clearly reflected in the matrix $K$, where Table I summarizes
how the scale dependence of $K_{11}$ and $\gamma_{12}$ depend on
disorder in 3D. In contrast, $K$ in Q1D is independent of
disorder. Our model recovers all the peculiarities of the 3D
localized regime: we found that the distribution $P(\ln g)$ is
narrower than in Q1D and possesses non-zero skewness.

Although we concentrated on 3D systems, general considerations
about the properties of the matrix $K$ in the insulating regime
should be valid in any dimension $d> 2$. In particular, the
assumption that $K_{11}$ is a nonzero $L$-independent constant in
the localized regime is correct independent of dimensionality.
Therefore, we expect that our theory of the insulating regime is
valid for any $d>2$. Consequently, the distribution $P(\ln g)$ is
not Gaussian in the localized regime in any $d>2$, although the
nature of the deviation from the Gaussian form might be
dimensionality dependent. In fact we expect $P(\ln g)$ to be
different from Gaussian distribution even in 2D \cite{markos2}.
This expectation is supported by Fig.~\ref{2D} which shows $P(\ln
g)$ for a 2D square system obtained numerically using
Eq.~\ref{AndHam}. As discussed in the paper, deviations from the
Gaussian form are due to the changes of the spectrum of parameters
$x$. Although the contribution of the first channel to the
conductance is dominant, higher channels do influence the
statistical properties of the smallest parameter, $x_1$. This
effect was probably not considered in previous analytical works
which predict Gaussian distributions of $\ln g$ in the insulating
regimes in dimensions $d=2+\epsilon$ \cite{altshuler}.

When applied to the critical regime, our theory recovers typical
properties of the conductance distribution, including the
non-analyticity of the distribution in the vicinity of $g=1$.
Although our present results are only qualitatively correct in the
critical regime, we believe that the method developed in the paper
represents a good starting point for further development of the
theory of Anderson transition.

\bigskip

KAM thanks U. Karlsruhe for support and hospitality during his
visit. PM thanks APVT, grant n. 51-021602 for financial support.
PW gratefully acknowledges support of a visit and hospitality at
the U. Florida as well as partial support by a Max-Planck Research
Award.


\begin{thebibliography}{40}
\bibitem{note} Note that for $|a-b|\gg L$, $\gamma_{ab}$ must decrease as
$L^{-2}$. Indeed, from (\ref{two}) we have $\sum_{b}K_{ab}=1$ so
that the sum of $L^2$ \textit{positive} parameters $\sum_b
\gamma_{ab}$ is $\propto L^0$ in the insulating regime. We ignore
this detail in our model.
\bibitem{noteG/g} In Ref. [\onlinecite{mmwk}] we used  $\Gamma/\gamma_{12}=4$.
This difference is not important, since, as we show in
Fig.~\ref{skew_gamma}, the actual value of $\Gamma/\gamma_{12}$
does not influence the value of the skewness  when both $\Gamma$
and $\gamma_{12}$ are small.
\end{thebibliography}
\end{document}